\documentclass[a4paper,11pt]{article}
\pdfoutput=1
\usepackage{jheppub}
\usepackage{slashed}
\usepackage{color}
\title{Interactions of a Stabilized Radion and Duality}
\author[a]{Zackaria Chacko,}
\author[a]{Rashmish K. Mishra,}
\author[b]{Daniel Stolarski,}
\author[a]{and Christopher B. Verhaaren}
\affiliation[a]{Maryland Center for Fundamental Physics, 
Department of Physics, \\ 
University of Maryland, College Park, MD 20742-4111}
\affiliation[b]{Theory Division, Physics Department, 
CERN, CH-1211 Geneva 23, Switzerland}

\preprint{CERN-PH-TH-2014-225}

\abstract
{
We determine the couplings of the graviscalar radion in Randall-Sundrum 
models to Standard Model fields propagating in the bulk of the space, 
taking into account effects arising from the dynamics of the 
Goldberger-Wise scalar that stabilizes the size of the extra dimension. 
The leading corrections to the radion couplings are shown to arise from 
direct contact interactions between the Goldberger-Wise scalar and the 
Standard Model fields. We obtain a detailed interpretation of the 
results in terms of the holographic dual of the radion, the dilaton. In 
doing so, we determine how the familiar identification of the parameters 
on the two sides of the AdS/CFT correspondence is modified in the 
presence of couplings of the bulk Standard Model fields to the 
Goldberger-Wise scalar. We find that corrections to the form of the 
dilaton couplings from effects associated with the stabilization of the 
extra dimension are suppressed by the square of the ratio of the dilaton 
mass to the Kaluza-Klein scale, in good agreement with results from the 
CFT side of the correspondence.
}
\begin{document}
\maketitle
\flushbottom

\section{Introduction}
\label{sec:intro}

The unambiguous discovery of a new scalar resonance with the properties 
expected of the Standard Model (SM) Higgs represents a milestone in the 
history of elementary particle physics. A careful study of the 
properties of this Higgs particle is expected to shed light on the 
dynamics that drives electroweak symmetry breaking. At present, an 
important open question is whether this state is an elementary particle, 
or a composite made up of more fundamental constituents held together by 
some form of new strong dynamics. Compositeness of the Higgs would allow 
a simple resolution of the hierarchy problem, provided the new strong 
dynamics kicks in at energies close to the weak scale, and therefore 
constitutes a very compelling theoretical possibility. However, the 
generation of fermion masses in composite Higgs scenarios is a 
challenge. The simplest models involve new sources of flavor violation 
close to the weak scale and are therefore disfavored by experiment.

An interesting class of composite Higgs models that can resolve this 
flavor problem are those where the new strong dynamics is conformal in 
the ultraviolet (UV). Strong conformal dynamics allows the flavor scale 
in these theories to be well separated from the weak scale, allowing the 
stringent experimental limits on flavor changing neutral currents to be 
satisfied. This scenario is closely related to earlier proposals for 
suppressing flavor violation in technicolor models~\cite{Luty:2004ye}, 
(see also~\cite{Holdom:1984sk, Appelquist:1986an, Yamawaki:1985zg, 
Appelquist:1986tr}). In this class of theories the conformal symmetry is 
spontaneously broken at low energies. As a consequence, if the conformal 
symmetry were exact, the low energy spectrum would contain a massless 
Nambu-Goldstone boson (NGB), the 
dilaton~\cite{Salam:1970qk,Isham:1970gz,Zumino:1970ab,Ellis:1970yd,Ellis:1971sa}. 
In this limit the form of the dilaton couplings to the SM fields can be 
completely determined from the requirement that the conformal symmetry 
be realized nonlinearly. 

In the theories of phenomenological interest, however, the conformal 
symmetry is only approximate. It is explicitly violated by operators 
that are small in the UV but grow large in the infrared (IR), thereby 
driving the breaking of conformal symmetry. Provided the operator 
primarily responsible for this breaking has a scaling dimension close to 
marginal, the theory can remain approximately conformal for enough 
decades in scale for the flavor problem to be addressed. However, as a 
consequence of the explicit breaking, the dilaton is not massless and 
its couplings receive corrections. It is important, therefore, to 
understand the exact circumstances under which the dilaton can remain 
light and to determine the size and form of the corrections to its 
couplings.

Recently, several authors have studied the conditions under which the 
low energy spectrum contains a light 
dilaton~\cite{Chacko:2012sy,Bellazzini:2012vz,Coradeschi:2013gda}.
 The general picture that has emerged 
is that if the operator $\mathcal{O}$ primarily responsible for the 
breaking of conformal symmetry is close to marginal at the breaking 
scale, the mass of the dilaton can naturally lie below the scale of the 
strong dynamics. This result is explained by the fact that the extent 
of explicit conformal symmetry violation at the breaking scale depends 
not just on the size of the deformation associated with $\mathcal{O}$, 
but also on the deviation from marginality of the operator 
$\mathcal{O}$. In particular, the theory will retain an approximate 
conformal symmetry if the operator $\mathcal{O}$ is very close to 
marginal, independent of the size of the deformation. In such a 
scenario, even if the deformation is large, the dilaton can naturally
be light provided $\mathcal{O}$ is close to marginal at the breaking 
scale. Unfortunately, unless the theory possesses some special feature, 
this condition is not expected to be satisfied and the dilaton is not 
light. The underlying reason for this is that, even if the operator 
$\mathcal{O}$ that drives the breaking of conformal symmetry is indeed 
close to marginal in the UV, as in the theories of phenomenological 
interest, its scaling behavior is expected to receive big corrections 
when the deformation grows large. Therefore, in general $\mathcal{O}$ 
does not remain marginal near the breaking scale where the deformation 
is large. As a consequence, the presence of a light dilaton in the 
spectrum is not a robust prediction of the class of theories of 
interest for electroweak symmetry breaking.

One special class of theories where the dilaton can naturally remain light 
are those which possess not just a single isolated fixed point, but an 
entire line of fixed (or quasifixed) points. This feature, which is 
quite common in supersymmetric theories, allows the deformation to remain 
marginal at the breaking scale. Other constructions which admit the 
possibility of a naturally light dilaton are gauge theories that lie near the edge of the conformal window~\cite{Appelquist:2010gy}. One scenario which allows the spectrum of light states to contain a dilaton, albeit at the expense of mild tuning, arises if the breaking of conformal symmetry occurs before the deformation associated with $\mathcal{O}$ reaches its natural strong coupling value. In this limit, because the size of the deformation is small, the corresponding corrections to the scaling behavior of the operator $\mathcal{O}$ at the breaking scale are also suppressed, allowing it to remain close to marginal. Then, the limited extent to which conformal symmetry is violated allows the dilaton to remain light. In general, however, the conformal symmetry is not expected to break until the deformation becomes large, so this scenario is associated with tuning. This tuning is mild, however, scaling only linearly with the mass of the dilaton \cite{Chacko:2012sy,Bellazzini:2012vz}. Therefore, the presence in the low energy spectrum of a dilaton with a mass just a factor of a few below the compositeness scale is associated with only modest tuning. From this discussion we see that a light dilaton can arise in several different realistic scenarios, and therefore the dynamics of theories with a light dilaton 
remains a problem of phenomenological interest.{\footnote{String motivated constructions that can give rise to a light dilaton have been considered, for example, in~\cite{Elander:2009pk,Elander:2012yh,Evans:2013vca,Elander:2014ola}.}

The form of the dilaton couplings to the SM states has been determined 
in the limit that effects that explicitly violate conformal symmetry are 
neglected. Both the case when the SM matter and gauge fields are 
composites emerging from the strong 
dynamics~\cite{Goldberger:2007zk,Fan:2008jk}, and the case when they are 
external elementary states~\cite{Vecchi:2010gj},\cite{Chacko:2012sy, 
Bellazzini:2012vz}, have been studied. Corrections to the form of the 
dilaton couplings arising from explicit conformal symmetry violating 
effects have also been studied~\cite{Chacko:2012sy}, and found to scale 
as the square of the ratio of the mass of the dilaton to the strong 
coupling scale. A physical understanding of this result may be obtained 
by noting that in the theories of interest with a light dilaton, the 
operator $\mathcal{O}$ is close to marginal at the breaking scale, even 
though the deformation associated with $\mathcal{O}$ may be large. If 
$\mathcal{O}$ were exactly marginal the conformal symmetry would be 
exact, and independently of the size of the deformation, the dilaton 
couplings would be of the form dictated by nonlinearly realized 
conformal invariance. In this limit, the corrections to the dilaton 
couplings that arise from the deformation do not, in general, vanish. 
However, these effects can be exactly absorbed into corresponding 
changes in the low energy parameters, leaving the form of the 
interactions unchanged. The size of the corrections to the form of the 
dilaton couplings is therefore dictated not just by the size of the 
deformation, but also by the deviation from marginality of the operator 
$\mathcal{O}$ at the breaking scale. However, as noted above, it is 
precisely these two effects that also determine the dilaton mass. Therefore, 
the size of the corrections to the form of the dilaton couplings is 
correlated with the mass of the dilaton. These corrections are therefore 
small and under good theoretical control if the dilaton is light. If, 
however, the deformation is large and the scaling behavior of the 
operator $\mathcal{O}$ deviates significantly from marginality, the 
dilaton mass is raised to the strong coupling scale, and the corrections 
to the form of the dilaton couplings become of order one.

The AdS/CFT 
correspondence~\cite{Maldacena:1997re,Gubser:1998bc,Witten:1998qj,Klebanov:1999tb} 
relates theories of strong conformal dynamics to theories of gravity in 
higher dimensions. Theories of phenomenological interest where the 
strong conformal dynamics is spontaneously broken giving rise to a 
composite Higgs are dual~\cite{ArkaniHamed:2000ds,Rattazzi:2000hs} to 
Randall-Sundrum (RS) models~\cite{Randall:1999ee} where the extra 
dimension is negatively curved and finite, with a brane at either end. 
In this correspondence, the dilaton is dual to the radion, the 
excitation corresponding to fluctuations in the size of the extra 
dimension~\cite{ArkaniHamed:2000ds,Rattazzi:2000hs}. In the original RS 
model, the hierarchy between the Planck and weak scales depends on the 
brane spacing, which is a free parameter. In this limit the radion is 
massless. The brane spacing, and the associated Planck-weak hierarchy, 
can be stabilized using the Goldberger-Wise (GW) 
mechanism~\cite{Goldberger:1999uk}. This mechanism introduces a bulk 
scalar field $\Phi$ which is sourced on the two branes, and has a 
potential in the bulk. It therefore acquires a vacuum expectation value 
(VEV) which varies as a function of position in the extra dimension, 
and contributes to the vacuum energy. Consequently, the brane spacing 
is stabilized and the radion acquires a mass. Since the RS model is one 
of the most promising candidates for physics beyond the SM, it is 
important to obtain an understanding of the mass and couplings of the 
radion in this framework.

By holography the GW scalar $\Phi$ is dual to the CFT operator 
$\mathcal{O}$, whose dynamics drives the breaking of conformal symmetry. 
Sourcing the GW field corresponds to a deformation of the CFT by this 
operator, with the VEV of the GW field corresponding to the size of the 
deformation. The bulk mass term for the GW field is related to the 
scaling dimension of $\mathcal{O}$, while the self-interaction terms in 
the bulk potential for $\Phi$ correspond to corrections to the scaling 
behavior of $\mathcal{O}$ that are important when the deformation grows 
large.

The conditions under which the low energy spectrum of the RS model 
contains a light radion after stabilization have been studied, and 
found to agree with the results for the dilaton from the CFT side of the 
correspondence~\cite{Chacko:2013dra,Bellazzini:2013fga,Coradeschi:2013gda}. 
The desired large hierarchy of scales can naturally arise if the mass 
term for the GW scalar is small. This corresponds to the scaling 
dimension of the dual operator $\mathcal{O}$ being close to marginal. 
However, for the spectrum to naturally contain a light dilaton, the 
coefficients of the self-interaction terms for the GW scalar must also 
lie below their natural strong coupling values. From the dual 
perspective this ensures that the corrections to the scaling behavior 
of $\mathcal{O}$ from the deformation remain small, even when the deformation itself is large, so that $\mathcal{O}$ remains close to marginal at the breaking scale. However, unless the 5D 
construction possesses some special feature, in general the 
self-interaction terms are not small and this condition is not 
satisfied. Therefore, the presence of a light radion in the low energy 
spectrum below the Kaluza-Klein (KK) scale is not a robust feature of 
RS models \cite{Chacko:2013dra}.

One special class of theories where the radion can naturally remain 
light are those where the GW scalar arises as the pseudo-Nambu Goldstone 
boson (pNGB) of an approximate global symmetry. In this case the mass and 
self-interaction terms in the potential for the GW scalar can naturally be small, thereby allowing the 
radion mass to lie below the KK scale. Several authors have considered this limit and found that the radion is indeed light, its mass scaling as the mass of the GW scalar~\cite{Konstandin:2010cd,Eshel:2011wz},\cite{Chacko:2013dra}. Careful studies have shown that the inclusion of gravitational backreaction does not alter this conclusion~\cite{Coradeschi:2013gda,Bellazzini:2013fga,Megias:2014iwa,Cox:2014zea}. This corresponds in the dual theory to the case when the CFT possesses a line of quasifixed points. An alternative scenario which allows the spectrum of light states to contain a radion, albeit at the expense of mild tuning, arises if, 
after stabilization, the VEV of the GW scalar in the neighborhood of the 
IR brane lies below its natural strong coupling value. This is dual to the 4D operator corresponding to $\mathcal{O}$ being below its strong coupling value at the breaking scale. In this limit, the overall contribution of the GW field to the potential for the radion and the effects of the self-interaction terms are both suppressed, allowing the radion to remain light. Although such a scenario is associated with tuning, the tuning is mild, scaling only linearly with 
the ratio of the mass of the radion to the KK scale \cite{Chacko:2013dra}.

Since the radion is the graviscalar excitation of the 
metric~\cite{Randall:1999ee}, the form of its interactions follows from 
general covariance~\cite{Goldberger:1999uk}. The radion couplings to SM 
fields have been determined, both in the case of brane-localized 
matter~\cite{Csaki:1999mp,Goldberger:1999un,Giudice:2000av,Csaki:2000zn}, 
and in the case of matter in the bulk~\cite{Rizzo:2002pq,Csaki:2007ns}. 
The dynamics associated with stabilization of the extra dimension leads 
to corrections to these couplings. Previous work to determine the form 
of these corrections was restricted to the technically simpler case of 
brane-localized fields~\cite{Chacko:2013dra}. In the dual picture, this 
corresponds to the case when all the SM fields are composites of the 
strong dynamics. The results obtained are in good agreement with those 
from the CFT side of the correspondence. The goal of this paper is to 
extend this analysis to the case when the SM matter and gauge fields 
reside in the bulk of the space. This scenario, which admits an elegant 
solution to the SM flavor 
problem~\cite{ArkaniHamed:1999dc,Grossman:1999ra,Gherghetta:2000qt,Agashe:2004cp}, 
corresponds in the dual picture to the SM fermions arising as partial 
composites of elementary particles and CFT states~\cite{Kaplan:1991dc}.

In what follows, we consider a scenario where the SM gauge bosons and 
fermions propagate in the bulk of the RS geometry, but the Higgs is 
localized to the IR brane. We stabilize the brane spacing by employing a 
GW scalar $\Phi$ that is sourced on the branes and determine the radion 
couplings to the bulk SM fields. This construction allows direct 
couplings of the GW scalar to SM fields in the bulk. To leading order in 
$\Phi$, these couplings take the schematic form
 \begin{eqnarray}
\sqrt{\left|G\right|}\:\mathcal{O}_{\text{SM}}\:\Phi \:.
\label{eq:bulk-op-phi-in-bulk}
 \end{eqnarray}
 Here $G$ is the determinant of the 5D RS metric, and 
$\mathcal{O}_{\text{SM}}$ is a gauge invariant operator composed of bulk 
SM fields. Brane localized interactions between the GW scalar and the SM 
fields are also expected to be present. The operators in 
\eqref{eq:bulk-op-phi-in-bulk} affect the masses and interactions of the 
fields in the low energy effective theory. We find that the leading 
corrections to the radion couplings to the SM fields arise from such 
terms, and perform a careful calculation to determine their effects. One 
might expect that the effects of the stabilization mechanism on the 
radion profile would lead to corrections to the radion couplings, even 
in the absence of direct couplings of the GW scalar to the SM particles. 
However, we show in Appendix \ref{app:mixing} that these effects are much smaller than the corrections obtained from operators of the form \eqref{eq:bulk-op-phi-in-bulk}.

We obtain a detailed interpretation of our results in terms of the 
holographic dual of the radion, the dilaton. In doing so, it is 
important to take into account the fact that the familiar identification 
of the parameters on the two sides of the AdS/CFT correspondence is 
modified in the presence of couplings of the bulk SM fields to the GW 
scalar. This is because one class of corrections to the radion couplings 
can be completely absorbed into changes in the parameters of the dual 
theory, and do not affect the form of the dilaton interactions. As in 
the case of brane-localized SM fields, we find that all corrections to 
the form of the dilaton couplings are suppressed by the square of the 
ratio of the dilaton mass to the KK scale, in good agreement with 
results from the CFT side of the correspondence.

These results have implications for phenomenological studies of the 
radion. Several authors have investigated the possibility that the 
resonance observed at 125 GeV is not the SM Higgs, but a dilaton/radion, 
for example~\cite{Cheung:2011nv,Matsuzaki:2012gd,Grzadkowski:2012ng,Matsuzaki:2012vc, 
Matsuzaki:2012mk,Elander:2012fk,Matsuzaki:2012xx,Chacko:2012vm}. Studies have also been performed using LHC data to place limits on the mass of the radion in RS models~\cite{Cho:2013mva,Desai:2013pga,Cao:2013cfa,Jung:2014zga,Boos:2014xha}, and investigating the prospects for detecting the radion at the LHC~\cite{Bhattacharya:2014wha} and future 
colliders~\cite{Cho:2014aka}. The dilaton has been investigated as a possible mediator of the interactions of dark matter with the SM \cite{Bai:2009ms,Agashe:2009ja,Blum:2014jca,Efrati:2014aea}. It has been shown that in certain theories the presence of a light radion can help explain the baryon asymmetry~\cite{Servant:2014bla}. In all these cases, an understanding of the size of the corrections to the radion couplings is necessary to understand the robustness of the conclusions.

The outline of this paper is as follows. In Sec.~\ref{sec:radion-stabilization} we provide the details of the GW mechanism that stabilizes the extra dimension and results in the radion acquiring a mass. We also explain the origin of the corrections to the radion couplings. In subsequent sections, we consider in turn the massless gauge bosons, massive gauge bosons, and fermions of the SM. For 
each case we determine the radion couplings and interpret the results 
from a holographic point of view. Details of the calculation are 
presented in the appendices.

\section{Radion Dynamics}
\label{sec:radion-stabilization}

In this section, we outline the steps involved in obtaining the mass and 
couplings of the radion in the presence of the GW mechanism. The 
discussion in this section closely follows~\cite{Chacko:2013dra}, and 
only the most relevant features are presented here. We begin with the 
5D action for the RS model in the absence of stabilization,
 \begin{eqnarray}
\mathcal{S}
=\int d^4x\:d\theta
\left[
\sqrt{G}\left(-2M_5^3\mathcal{R}[G]-\Lambda_b\right)
-\sqrt{-G_{\text{UV}}}\delta(\theta)T_{\text{UV}}
-\sqrt{-G_{\text{IR}}}\delta(\theta-\pi)T_{\text{IR}}
\right]\:.
\label{eq:grav-lag}
\end{eqnarray}
Here $M_5$ is the 5D Planck mass, $\Lambda_b$ is the bulk 
cosmological constant, and $T_{\text{UV}}$, $T_{\text{IR}}$ are the brane tensions on the UV and IR branes. The extra dimensional coordinate $\theta$ is compactified over $\mathcal{S}^1$ and the region $[-\pi,0)$ is identified with $[0,\pi)$ by a $\mathcal{Z}_2$ symmetry. The locations $\theta=0, \, \pi$ correspond to the locations of the UV and IR branes respectively. The static metric\footnote{We use (+ -- -- -- --) signature.} that describes the geometry of the two brane RS model is obtained as the solution to the 5D Einstein equations and can be written as
 \begin{eqnarray}
ds^2 =  e^{-2kr_c|\theta|}\eta_{\mu\nu}dx^\mu dx^\nu-r_c^2d\theta^2\qquad\qquad -\pi\leq\theta<\pi\:.
\label{eq:RS-metric}
 \end{eqnarray}
Here $k$ is the inverse curvature and the constant $r_c$ is 
proportional to the distance between the two branes. The parameter $k$ 
is related to the bulk cosmological constant and 5D Planck scale by
 \begin{eqnarray}
\Lambda_b = -24M_5^3k^2\:.
\label{eq:def_k}
 \end{eqnarray}
 A condition for the existence of a static solution of this form is that 
the brane and bulk cosmological constants satisfy the relation 
$\Lambda_b=k T_{\text{IR}}=-k T_{\text{UV}}$. The value of $r_c$ is a free parameter, corresponding to the fact that the brane spacing in the RS solution is undetermined. When we include a stabilization mechanism for the size
of the extra dimension, we can detune the tension of the IR brane away from the RS value and still obtain a static solution~\cite{Goldberger:1999uk}. 

When fluctuations about this background are considered, the low energy 
spectrum is found to contain, in addition to the massless 4D graviton, 
a massless radion field associated with the fluctuations of the brane 
spacing. To obtain the low energy effective theory for the light fields, 
we replace in the 5D metric $\eta_{\mu\nu}$ by the dynamical field 
$g_{\mu\nu}(x)$ and $r_c$ by $r(x)$. These fields are identified with 
the 4D graviton and the radion fields respectively. The metric is then 
substituted back into the 5D action. After integrating over the extra 
dimension, the resulting 4D action describes the low energy effective 
theory of the graviton and the radion,
\begin{eqnarray}
\mathcal{S} &=& \int d^4x\: \sqrt{-g}\:\left(\frac{2M_5^3}{k}\mathcal{R}[g_{\mu\nu}]+\frac12 \partial_\mu\varphi\partial^\mu\varphi\right)\:.
\label{eq:4D-rad-action}
\end{eqnarray} 
Here $\varphi$ represents the canonically normalized radion field and 
is related to $r(x)$ by
\begin{eqnarray}
\varphi(x) &=& \sqrt{\frac{24M_5^3}{k}}e^{-k \pi r(x)}\:.
\label{eq:rad2r}
\end{eqnarray}

The absence of a potential for $\varphi$ reflects the fact that the 
value of $r_c$ is undetermined. Stabilization of the extra dimension is 
accomplished by adding a bulk GW scalar $\Phi$ to the theory. This 
scalar acquires a $\theta$ dependent VEV, $\widehat{\Phi}(\theta)$, from potentials on the branes and in the bulk. Its VEV is also a function of $r_c$. The Lagrangian for the 4D effective theory, including the contribution of the GW field, may be obtained in the same manner as before. Specifically, after replacing $r_c$ by $r(x)$, $\widehat{\Phi}$ is substituted back into the action and the integration over the extra dimension is performed. The resulting 4D action includes the contribution of the GW scalar to the low energy theory. This effect generates a potential for $\varphi$ that, 
when minimized, fixes $r_c$ and gives mass to the physical radion field.

To understand this in more detail, consider the action for the GW scalar,
 \begin{align}
\mathcal{S}_{\text{GW}}
&=\int d^4xd\theta
\left[
\sqrt{G}\left(
\frac12G^{AB}\partial_A\Phi\partial_B\Phi
-V_b(\Phi)
\right)
-\sum_{i=\text{IR,UV}}\delta(\theta-\theta_i)\sqrt{-G_i}V_i(\Phi)
\right].
\label{eq:gw-action}
\end{align}
Here $V_{\text{UV}}$ and $V_{\text{IR}}$ are the potentials on the UV and IR branes and $V_b$ is the potential in the bulk. For simplicity, we choose to work 
with a linear potential on the IR brane,
 \begin{eqnarray}
V_{\text{IR}} &=& 2\alpha k^{5/2} \Phi \, .
\label{eq:gw-vis-brane-pot}
 \end{eqnarray}
This is a consistent choice if $\Phi$ is not charged under any 
symmetries, and the qualitative features of our results do not depend on 
the specific form of this potential. On the UV brane we do not 
specify a form of the potential but require that the value of $\Phi$ is 
$k^{3/2}v$. This requirement is satisfied for many choices of potentials 
including the one considered in the original GW 
proposal~\cite{Goldberger:1999uk}. In order to generate a sizable
hierarchy, the size of the extra dimension must be large in units of the 
curvature. To accomplish this, we require that $v$ be somewhat smaller 
than its natural strong coupling value. 

The bulk potential for $\Phi$ is of the general form
 \begin{eqnarray}
V_b(\Phi) &=& \frac{1}{2} m^2\Phi^2+\frac{1}{3!}\eta\:\Phi^3+\frac{1}{4!}\zeta\:\Phi^4+\ldots\:.
\label{eq:bulk-pot}
 \end{eqnarray}
The bulk mass parameter $m^2$ of the GW scalar must be small in units 
of the inverse curvature $k$ to address the large Planck-weak hierarchy. 
However, there are no such requirements on the cubic and higher order 
terms. This can be understood from the holographic perspective. AdS/CFT 
relates the extra dimensional coordinate $\theta$ to the renormalization 
scale $\mu$ in the dual 4D theory, ${\rm{log}}(k/\mu) \sim k r_c 
\theta$. The duality also relates the value of the GW field 
$\Phi(kr_c\theta)$ at any point $\theta$ in the bulk to the size of the 
coefficient of the operator that deforms the dual CFT at the 
corresponding scale $\mu$. Therefore, requiring that the value of $v$ on 
the UV brane be small corresponds to requiring that the size of the 
deformation be small at high scales $\mu \sim k$. Then, if the bulk mass 
term is also small, the initial growth in the value of $\Phi$ is slow, 
allowing a large hierarchy to develop. In the dual picture, the mass of 
the GW scalar is related to the scaling dimension of the dual operator. 
A massless scalar corresponds to an exactly marginal deformation, while 
a negative mass squared for $\Phi$ corresponds to a relevant operator in 
the dual CFT. Note that a negative mass squared for $\Phi$ in AdS space is free from any instabilities for $|m^2| \leq 4 k^2$~\cite{Breitenlohner:1982jf} and corresponds to the scaling dimension of the operator in the dual theory being relevant. If the mass term is small and negative, the deformation is 
relevant, but close to marginal. This allows the coefficient of this 
relevant operator to start at a small value at high energies and grow 
slowly, leading to a large hierarchy before it eventually becomes strong 
enough to trigger breaking of the conformal symmetry. This is the 
scenario we shall focus on.

Higher order terms in the bulk potential correspond to corrections to 
the scaling behavior of the dual operator that become important when the 
deformation grows large. As the value of $\Phi$ becomes large close to 
the IR brane, the higher order interaction terms are expected to 
dominate over the suppressed mass term unless they are also small from 
symmetry considerations, as in the case where $\Phi$ is a pNGB. For 
simplicity, we consider a scenario where the detuning of the IR brane 
tension away from the pure RS solution is slightly below its natural 
strong coupling value by a factor that could be as small as a 
few~\cite{Chacko:2013dra}. This allows the extra dimension to be 
stabilized when the VEV of the GW field in the neighborhood of the IR 
brane is also slightly below its natural strong coupling value. This 
limit captures the qualitative features we are interested in, but allows 
the gravitational backreaction to be neglected. In the dual picture, 
this corresponds to the assumption that the breaking of conformal 
symmetry is triggered when the deformation is still slightly below its 
strong coupling value. For this choice of parameters the cubic 
self-interaction term in the GW potential is expected to dominate over 
the other higher order terms. Therefore, in what follows, we keep only 
the mass and cubic terms in the bulk potential for $\Phi$ and neglect 
the higher order terms.

This limit also allows an approximate solution to the equations of motion for $\widehat{\Phi}$. The equations and the boundary conditions are 
given by
 \begin{eqnarray} 
&& 
\partial_\theta^2\widehat{\Phi}-4kr_c\partial_\theta\widehat{\Phi} -r_c^2m^2\widehat{\Phi}-r_c^2\frac{\eta}{2}\widehat{\Phi}^2=0 \nonumber \\ && 
\theta=0\qquad:\qquad\widehat{\Phi}=k^{3/2}v \nonumber \\ && 
\theta=\pi\qquad:\qquad\partial_\theta\widehat{\Phi}=-\alpha k^{3/2} kr_c\:. 
\end{eqnarray} 
 For notational simplicity, we trade the parameters $m$ and $\eta$ in 
the bulk potential of $\Phi$ for $\epsilon$ and $\xi$, which are given 
by
 \begin{eqnarray} 
\epsilon \equiv \frac{m^2}{4k^2}\:\:,\:\: \xi \equiv \frac{\eta v}{8\sqrt{k}}\:. 
\label{eq:eps-and-xi} 
 \end{eqnarray}
In the limit that the hierarchy is large, $kr_c\gg1$, the solution of 
this equation exhibits boundary layer structure~\cite{Chacko:2013dra}. 
This allows an approximate solution to be obtained using boundary layer 
analysis. Using these methods, the solution for $\widehat{\Phi}$ is found to be of the form
 \begin{align}
\widehat{\Phi}(\theta)
&=-\frac{k^{3/2}\alpha} {4}e^{-4kr_c(\pi - \theta)}+\widehat{\Phi}_{\text{OR}}(kr_c\theta)\nonumber\\
&=-\frac{k^{3/2}\alpha} {4}e^{-4kr_c(\pi - \theta)}+\frac{k^{3/2}v e^{-\epsilon kr_c\theta}}
{1+\xi\left(1-e^{-\epsilon kr_c\theta}\right)/\epsilon} \; .
\label{eq:phi-soln-massANDcubic}
\end{align}
While we have been specific about the $kr_c\theta$ dependence of 
$\widehat{\Phi}_{\text{OR}}$ in the above expression, we shall usually just write $\widehat{\Phi}_{\text{OR}}(\theta)$. Several features of this classical solution are now apparent:
\begin{itemize}
	
\item 
 The boundary region term, proportional to $\alpha$, is exponentially suppressed as long as one is away from the region $\pi - \theta \lesssim \epsilon$. This region is the ``boundary layer'' where the $\alpha$ term becomes important in the classical solution and $\widehat{\Phi}$ changes very quickly. In the dual 4D theory, this region corresponds to the energy scales at which the phase transition associated with the spontaneous breaking of conformal 
symmetry occurs.
	
\item 
 The second term, or outer region solution $\widehat{\Phi}_{\text{OR}}(\theta)$, depends on the mass $m^2$ and the cubic coupling $\eta$. If we make the cubic coupling small by setting $\xi$ to zero and work in the limit $\epsilon <0,|\epsilon| \ll 1$, $\widehat{\Phi}$ grows slowly with $\theta$, allowing a large hierarchy to be realized. As discussed above, a negative mass squared corresponds to the operator in the dual theory having a relevant scaling dimension.
	
\item 
 In the presence of a nonzero $\xi$ in $\widehat{\Phi}_{\text{OR}}$, the VEV again starts small and grows slowly, its growth controlled by the 
small parameter $\epsilon$. For $\theta$ away from $\pi$, the term 
multiplying $\xi$ is small and shields the effect of a nonzero $\xi$. As 
$\theta$ approaches $\pi$, however, the presence of $\xi$ cannot be 
ignored. Choosing a negative $\xi$ (and equivalently $\eta$) leads to a 
faster growth of $\widehat{\Phi}$ as $\theta$ approaches $\pi$. The cubic term is dual to the leading correction to the scaling behavior of the dual operator. 

\end{itemize}

We see that the qualitative features of this classical solution can be 
understood from holography and allow us to identify the range and sign 
of parameters in the AdS side of the correspondence. A plot comparing the 
classical solutions in the presence and absence of the cubic term is 
shown in Fig.~\ref{fig:phi-soln-massANDcubic}. We see that in the 
presence of the additional cubic interaction, $\widehat{\Phi}$ starts out the same but then grows faster with increasing $\theta$. 

\begin{figure}[ht]
\centering
\includegraphics[width=0.7\textwidth]{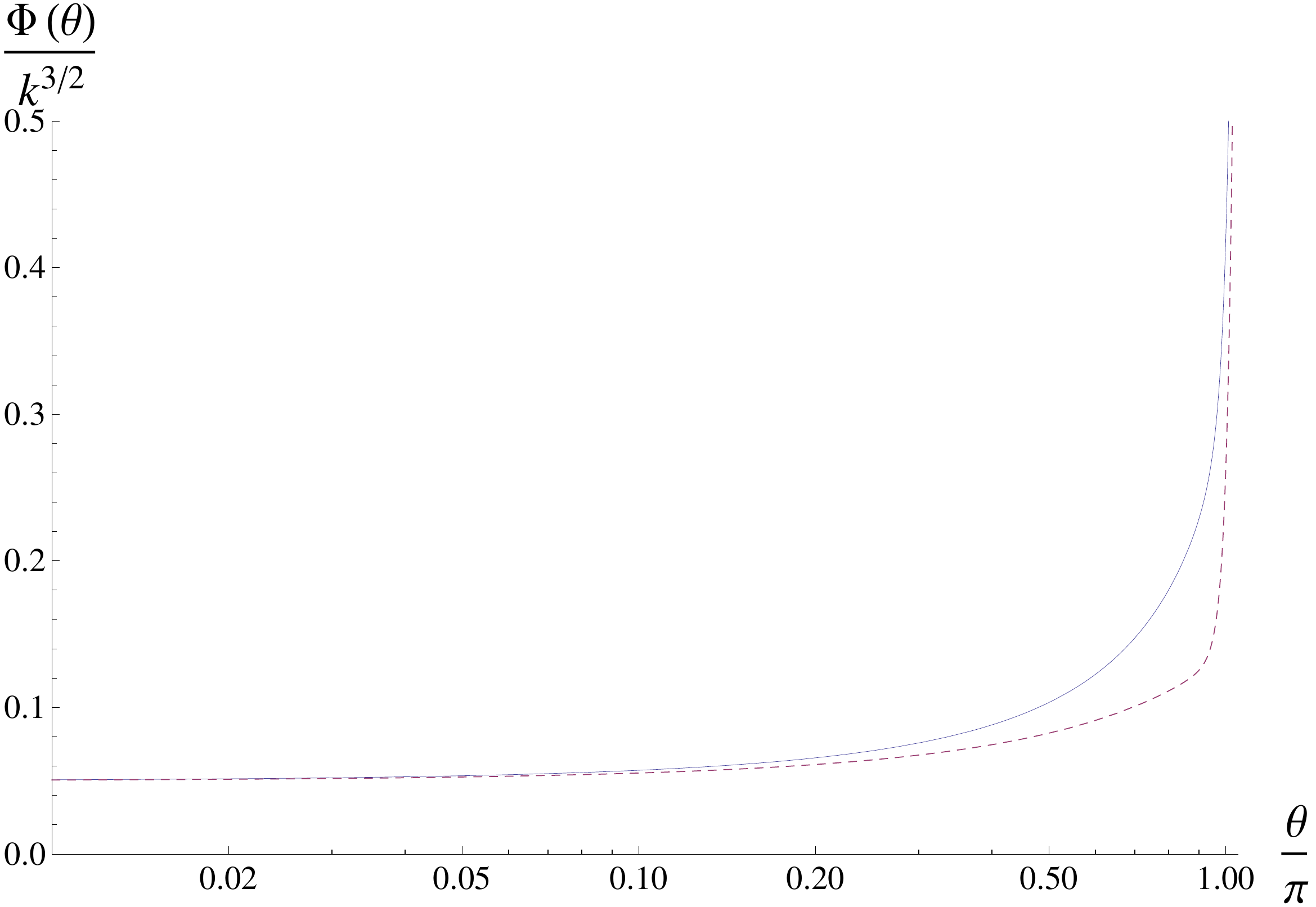}%
\caption{A comparison of the classical solution for $\widehat{\Phi}$ in the presence of an additional cubic bulk interaction term (solid line) compared to only having a bulk mass term (dashed), on a log scale for $\theta$. The choice of parameters are $\epsilon=-0.1$, $k\pi r_c=10$, $v=0.05$, $\alpha=-0.5$, $\xi=-0.03$. The effect of the cubic term is negligible near the $\theta=0$ boundary, but becomes important for larger $\theta$.}%
\label{fig:phi-soln-massANDcubic}%
\end{figure}

Once $r_c$ is made dynamical, $\widehat{\Phi}$ generates a contribution 
to the radion potential leading to a mass for the radion.{\footnote{The radion potential also receives contributions from the SM gauge bosons and fermions in the bulk through the Casimir effect~\cite{Goldberger:2000dv,Garriga:2002vf}. Since this contribution is loop suppressed, it is much smaller than the classical effect associated with the GW scalar, and can be neglected.}}
The dynamics associated with radion stabilization affects the couplings 
of the radion field. In general, the GW scalar has contact interactions 
with the SM fields. Once $\Phi$ acquires a VEV, these interactions alter 
the parameters in the low energy theory and correct the radion couplings 
to SM states. To understand schematically how these effects arise, 
consider the following operator which couples $\Phi$ to the SM fields,
 \begin{eqnarray}
\mathcal{L} \supset \sqrt{G}\:\mathcal{O}_{\rm SM}(x,\theta)\:\frac{\Phi(\theta)}{k^{3/2}}\:.
\label{eq:GW-SM-op}
\end{eqnarray}
 Here $\mathcal{O}_{\rm SM}$ is a gauge invariant operator made of SM 
fields. The operator in Eq.~\eqref{eq:GW-SM-op} is expected to be 
present in the absence of any symmetries that prohibit it. In the limit 
of small backreaction, $\widehat{\Phi}$ is slightly below its natural 
strong coupling value, so that this operator is expected to dominate 
over similar terms involving higher powers of $\widehat{\Phi}$. 
Replacing $r_c$ by $r(x)$ and working in terms of the canonical radion 
$\varphi$, $\widehat{\Phi}$ can be expanded as
 \begin{eqnarray}
\widehat{\Phi}(\theta, \varphi(x)) = \widehat{\Phi}(\theta, f) + \partial_\varphi \widehat{\Phi}(\theta,f)\:\left(\varphi-f\right)+\ldots\:,
\end{eqnarray}
 where $f=\left<\varphi\right>$. After substituting this expansion 
into Eq.~\eqref{eq:GW-SM-op}, performing a KK expansion and integrating over the extra dimension, we find that the first term in the $\widehat{\Phi}$ expansion can be absorbed into redefinitions of the parameters of the theory, while the second and subsequent terms generate corrections to the coupling of the radion to the zero modes of the SM fields contained in $\mathcal{O}_{\rm SM}$.

In the next few sections, we determine the couplings of the radion to
bulk fields such as SM gauge bosons and fermions, focusing on how
operators like Eq.~\eqref{eq:GW-SM-op} modify the leading order story. To write the corrections to radion couplings in a meaningful form, we define
$\frac{d\phantom{kr_c\theta}}{d(kr_c\theta)}\widehat{\Phi}\equiv \widehat{\Phi}'$.  Then, to leading order in $\widehat{\Phi}_{\text{OR}} '$ and $e^{-k\pi r_c}$, we find
\begin{equation}
\frac{m_\varphi^2}{\Lambda_{\text{IR}}^2}
= -\frac{\alpha k^3}{6 M_5^3} k^{-3/2}\widehat{\Phi}_{\text{OR}} '(kr_c\pi),
\label{eq:rad-genmass-relation}
\end{equation} 
where $m_\varphi$ is the mass of the radion and $\Lambda_{\text{IR}}\sim k e^{-k\pi r_c}$ is the KK scale. We note that in the two physical limits where the interactions are suppressed ($\eta\to 0$) or the mass is very small ($m^2 \to 0$), the expression for $\widehat{\Phi}_{\text{OR}}$ simplifies considerably,
\begin{eqnarray}
\widehat{\Phi}(\theta) = -\frac{k^{3/2}\alpha}{4}e^{4kr_c(\theta-\pi)}
+\left\{
\begin{array}{ll}
\displaystyle k^{3/2}v e^{-\epsilon kr_c\theta} &, \qquad \xi \rightarrow 0
\\
&
\\
\displaystyle
\frac{k^{3/2}v}
{1+\xi k r_c\theta} &, \qquad \;\epsilon \rightarrow 0
\end{array}
\right. ,
\end{eqnarray}
and Eq. \eqref{eq:rad-genmass-relation} can be simplified to
 \begin{eqnarray}
\eta &\rightarrow& 0\; : \qquad\frac{m_\varphi^2}{\Lambda_{\text{IR}}^2}
= \frac{\alpha k^3}{6 M_5^3} \epsilon\, v\, e^{-\epsilon k r_c \pi } 
\nonumber
\\
m^2 &\rightarrow& 0\; : \qquad \frac{m_\varphi^2}{\Lambda_{\text{IR}}^2}
= \frac{\alpha k^3}{6 M_5^3}
\frac{\xi v}{(1+\xi k\pi r_c)^2}
\:.
\label{eq:rad-mass-relations}
\end{eqnarray}
These results may be obtained by analyzing the minimization condition for the radion potential and the expressions for the mass of the radion in each 
case~\cite{Chacko:2013dra}.

\section{Massless Gauge Bosons}
\label{sec:radion2gauge-bosons-massless}

In this section, we determine the radion couplings to the massless gauge 
bosons of the SM, the photon and the gluon. We begin by considering the 
theory in the absence of a stabilization mechanism. The relevant part of 
the action is given by
 \begin{eqnarray}
\mathcal{S}
&=&
\int d^4x\;d\theta
\left[
-\frac{\delta(\theta)\sqrt{-G_{\text{UV}}}}{4g_{\text{UV}}^2}F^2
-\frac{\sqrt{G}}{4g_5^2}F^2
-\frac{\delta(\theta-\pi)\sqrt{-G_{\text{IR}}}}{4g_{\text{IR}}^2}F^2
\right]\, ,
\label{eq:gauge-kt-bulkbrane}
 \end{eqnarray}
where $F^2 = G^{MK}G^{NL}F_{MN}F_{KL}$ and $g_{\text{UV}}$, $g_5$, and $g_{\text{IR}}$ represent the gauge couplings on the UV brane, in the bulk, and on the IR brane.

After KK decomposition of the 5D action, we find that the spectrum of gauge bosons consists of a massless zero mode and heavier KK modes. The zero mode, 
which is identified with the corresponding massless gauge boson of the 
SM, has a flat profile in the extra dimension. To obtain the effective 
theory involving the massless mode, which we denote by $A_\mu(x)$, we 
simply replace $A_\mu(x,\theta)$ by $A_\mu(x)$ in the action and 
integrate over the extra dimension. Then the Lagrangian for the massless 
gauge bosons in the 4D effective theory takes the form
 \begin{equation}
-\frac{1}{4} \frac{1}{g_4^2} F_{\mu \nu} F^{\mu \nu} \:,
\label{kt}
 \end{equation}
 where the 4D gauge coupling $g_4$ at the KK mass scale is related to 
the underlying parameters of the 5D theory by
 \begin{eqnarray}
\frac{1}{g_4^2}=\frac{1}{g_{\text{UV}}^2}+\frac{2\pi r_c}{g_5^2}+\frac{1}{g_{\text{IR}}^2}\:.
\label{eq:gauge-coup}
 \end{eqnarray} 
 To obtain the coupling of the zero mode to the radion, we substitute 
the metric from Eq.~\eqref{eq:RS-metric} into Eq.~\eqref{eq:gauge-kt-bulkbrane} 
and promote $r_c$ to a dynamical field $r(x)$. Expressing the result in 
terms of the canonically normalized radion field $\varphi$ and expanding 
about its VEV $\left<\varphi\right>=f$, we obtain the coupling of the 
zero mode to the physical radion $\widetilde{\varphi}=\varphi-f$. The 
result, in a basis where the gauge kinetic term is normalized as in 
Eq.~\eqref{kt}, takes the form~\cite{Rizzo:2002pq,Csaki:2007ns}
 \begin{eqnarray} 
\frac{1}{2kg_5^2}\frac{\widetilde{\varphi}}{f}\:F_{\mu\nu}F^{\mu\nu}\; ,
\label{eq:zeromode-GB-rad-class} 
\end{eqnarray} 
 where indices are raised and lowered using the Minkowski metric 
$\eta^{\mu\nu}$. In contrast to the case of massless gauge bosons 
localized on the IR brane~\cite{Giudice:2000av}, we see that in 
this scenario the classical contribution to the coupling does not 
vanish. In Appendix~\ref{app:c} we estimate the natural size of the 
bulk gauge $g_5$ coupling in units of $k$. We find that $1/2kg_5^2$ is 
expected to be small, $1/2kg_5^2 \ll 1$. 

The one-loop quantum contribution to the radion coupling to the massless 
gauge bosons is also important, potentially comparable in size to the 
effect in Eq.~\eqref{eq:zeromode-GB-rad-class}. To determine this 
effect, note that the value of the 4D gauge coupling below the KK scale 
is in general a function of the background radion field. At low 
energies, the 4D gauge coupling satisfies a one-loop renormalization 
group (RG) equation of the form
 \begin{eqnarray}
\frac{d}{d\log \mu} \frac{1}{g^2(\mu)} &=& 
\frac{b_<}{8\pi^2}\:, \qquad \Lambda_{\rm IR} \geq\mu\geq0\:,
\label{eq:gaugeRG}
 \end{eqnarray}
where $\Lambda_{\rm IR}$ represents the cutoff of the 4D effective 
theory and scales with $r_c$ as
\begin{eqnarray}
\Lambda_{\rm IR} \sim m_{\text{KK}}\sim ke^{-k\pi r_c}\:.
 \end{eqnarray} 
 The value of the gauge coupling at the cutoff $g(\Lambda_{\rm IR})$ is 
identified with $g_4$ in Eq.~\eqref{eq:gauge-coup}. The quantity $b_<$ 
receives contribution from the particles in the spectrum below 
$\Lambda_{\rm IR}$ that run in the loops that renormalize the gauge 
coupling. We can solve Eq.~\eqref{eq:gaugeRG} to obtain the 4D gauge 
coupling at scales $\mu<\Lambda_{\rm IR}$:
 \begin{eqnarray}
\frac{1}{g^2(\mu)}= \frac{1}{g_4^2} 
- \frac{b_<}{8\pi^2}\log\left(\frac{\Lambda_{\rm IR}}{\mu}\right)\:.
\label{eq:gcoupling-low-scale}
 \end{eqnarray}
To compute the corresponding one-loop contribution to the radion-gauge 
boson vertex, we promote the parameter $r_c$ contained in $\Lambda_{\rm IR}$
in Eq.~\eqref{eq:gcoupling-low-scale} to a dynamical field and expand about 
its VEV. The kinetic term in the low energy theory
 \begin{eqnarray}
-\frac{1}{4g^2(\mu)}F_{\mu\nu} F^{\mu\nu}\:
\end{eqnarray}
 then generates a coupling to the normalized radion that is given by
 \begin{eqnarray}
\frac{b_<}{32\pi^2}\frac{\widetilde{\varphi}}{f}F_{\mu\nu}^2\:.
\label{eq:zeromode-GB-rad-loop}
 \end{eqnarray}
Combining this with~\eqref{eq:zeromode-GB-rad-class}, the full radion coupling is given by
 \begin{eqnarray}
\left(\frac{1}{2kg_5^2}+\frac{b_<}{32\pi^2}\right)\frac{\widetilde{\varphi}}{f}F_{\mu\nu} F^{\mu\nu} \:.
\label{eq:zeromode-GB-rad-full}
 \end{eqnarray}

To understand this result from a holographic point of view, recall that 
the AdS/CFT dictionary relates the bulk coordinate $\theta$ to the RG scale $\mu$ in the dual theory. The position of the UV brane corresponds to the cutoff $\Lambda_{\rm UV}\sim k$ of the CFT, while the position of the IR brane is associated with the scale $\Lambda_{\rm IR} \sim k e^{-k\pi r_c}$, where the CFT is spontaneously broken. Holography also relates a bulk gauge symmetry in the two brane AdS space to the weak gauging of a global symmetry in the dual CFT~\cite{ArkaniHamed:2000ds,Rattazzi:2000hs}. In general, this 
gauge coupling is expected to run with the RG scale:
 \begin{eqnarray}
\frac{d}{d\log \mu} \frac{1}{g^2(\mu)} &=&
\frac{b_>}{8\pi^2}\:, \qquad \Lambda_{\rm UV} \geq\mu\geq \Lambda_{\rm IR}\:.
\label{eq:running-gauge-coupling}
 \end{eqnarray}

To relate $b_>$ to the parameters of the dual AdS theory, we take the 
following approach. Consider moving the UV brane from $\theta = 0$ to an 
arbitrary point $\theta = \theta_0$ in the bulk. This corresponds to 
lowering the cutoff of the theory from $\Lambda_{\rm UV} \sim k$ to the 
scale $\Lambda_0$, given by
 \begin{eqnarray}
\Lambda_{\rm UV} \exp(- k\theta_0 r_c) = \Lambda_0 \; .
\label{eq:shifted-gauge-cutoff}
 \end{eqnarray} 
 The parameter $b_>$ can be determined by studying the corresponding 
change in the gauge coupling. We split the $\theta$ integral 
in the 5D action Eq.~\eqref{eq:gauge-kt-bulkbrane} into two parts, one 
from $0$ to $\theta_0$ and another from $\theta_0$ to $\pi$,
 \begin{eqnarray}
\mathcal{S}=\mathcal{S}_{\theta<\theta_0}+\mathcal{S}_{\theta>\theta_0}.
\label{eq:running-gauge-int}
 \end{eqnarray}
Substituting the zero mode back into the action, we evaluate the contribution to the $\theta$ integral from $\theta<\theta_0$ and match to the 
theory with the lower cutoff. This determines the correction to the 
brane localized kinetic term localized at $\theta_0$,
 \begin{eqnarray}
\frac{1}{g_{\text{UV}}^2(\theta_0)}=\frac{1}{g_{\text{UV}}^2} +\frac{2\theta_0r_c}{g_5^2}.\label{eq:betamore-begin}
 \end{eqnarray} 
 The effective 4D gauge coupling at the scale $\Lambda_0$, which we 
denote by $g_4(\Lambda_0)$, is equal to $g_{\text{UV}}(\theta_0)$ (up to small 
corrections of order $g_{\text{UV}}^2/kg_5^2$). This allows us to obtain the beta 
function at the scale $\Lambda_0$,
 \begin{eqnarray}
\frac{b_>}{8\pi^2}&\equiv&\frac{d}{d\log\Lambda_0}\frac{1}{g_{\text{UV}}^2(\Lambda_0)}
\nonumber\\
&=&-\frac{1}{kr_c} \frac{d}{d\theta_0}\frac{1}{g_{\text{UV}}^2(\theta_0)}
=-\frac{2}{kg_5^2} \;.
\label{eq:betamore-nointeraction}
\end{eqnarray}
 Notice that the expression for $b_>$ is independent of $\Lambda_0$. 
Using this, we can rewrite Eq.~\eqref{eq:zeromode-GB-rad-full} as\footnote{ In determining $b_>$ in Eqs. \eqref{eq:betamore-begin} and \eqref{eq:betamore-nointeraction}, we have not taken into account the quantum contributions from states localized on or toward the UV brane. However, because these states contribute equally to $b_<$ in Eq. \eqref{eq:gaugeRG}, their net contribution to $b_> - b_<$ in Eq. \eqref{eq:dilaton-beta} vanishes. As a consequence, they do not affect the final result. }
 \begin{eqnarray}
\frac{b_<-b_>}{32\pi^2}\frac{\widetilde{\varphi}}{f}\:F_{\mu\nu} F^{\mu\nu}\:.
\label{eq:dilaton-beta}
 \end{eqnarray} 
 This expression agrees with results obtained directly from the CFT side 
of the correspondence~\cite{Chacko:2012sy,Bellazzini:2012vz}.

We now include the effects of stabilization. In the dilaton case, the 
corrections to the form of Eq.~\eqref{eq:dilaton-beta} arising from the 
explicit breaking of the CFT are one-loop suppressed and scale as 
$m_\varphi^2/\Lambda_{\text{IR}}^2$, where $m_\varphi$ is now the dilaton mass and $\Lambda_{\text{IR}}$ is the cutoff of the effective theory where we expect 
composite states to appear. For the radion, the leading corrections arise from direct couplings of the GW scalar to gauge bosons in the bulk and on the branes. To leading order in $\Phi$, the effect is captured by
 \begin{eqnarray}
\mathcal{L}_{\text{int}}
&=&\frac{\Phi}{k^{3/2}}
\left[
-\beta_{\text{UV}}\frac{\delta(\theta)\sqrt{-G_{\text{UV}}}}{4g_{\text{UV}}^2}F^2
-\beta\frac{\sqrt{G}}{4g_5^2}F^2
-\beta_{\text{IR}}\frac{\delta(\theta-\pi)\sqrt{-G_{\text{IR}}}}{4g_{\text{IR}}^2}F^2
\right] \; .
\label{eq:GW-bulkGB-int}
 \end{eqnarray} 
 Here $\beta_{\text{UV}}, \,\beta$ and $\beta_{\text{IR}}$ are dimensionless numbers. When we replace $\Phi$ by its VEV and consider fluctuations of the radion about its background value, these interaction terms generate corrections 
to the 4D gauge coupling in the low energy effective theory, and to 
the radion coupling to the gauge bosons. Gauge invariance requires 
that the zero mode $A_\mu(x)$ continue to have a flat profile even in 
the presence of the $\Phi$ terms, but the relationship between the 4D 
gauge coupling $g_4$ and the underlying 5D parameters 
of Eq.~\eqref{eq:gauge-coup} now becomes
 \begin{eqnarray}
\frac{1}{g_4^2}=\frac{1}{g_{\text{UV}}^2}\left(1+\beta_{\text{UV}} \frac{\widehat{\Phi}(0)} {k^{3/2}}\right)+\frac{2r_c}{g_5^2}\int_0^\pi d\theta \left(1+\beta \frac{\widehat{\Phi}(\theta)}{k^{3/2}}\right)
+\frac{1}{g_{\text{IR}}^2}\left(1+\beta_{\text{IR}} \frac{\widehat{\Phi}(\pi)}{k^{3/2}}\right) \; .
\label{eq:gauge-coup-withGWscalar}
 \end{eqnarray}

When determining the corrections to the couplings of the radion arising 
from the GW field it is useful to employ the identity
 \begin{equation}
\left.\widehat{\Phi}(\theta)\right|_{r=r_c+\delta r}=\left.\widehat{\Phi}(\theta)\right|_{r=r_c}+\delta rk^{5/2}\alpha(\pi-\theta)e^{-4kr_c(\pi-\theta)}+\delta r\theta k\widehat{\Phi} '_{\text{OR}}(\theta kr_c)\,,
 \end{equation}
 with $ \widehat{\Phi}_{\rm OR} ' \equiv\frac{d\phantom{kr_c\theta}}{d(kr_c\theta)}\widehat{\Phi}_{\rm OR}$. After integrating over the extra dimension the contribution to the radion coupling from classical effects is obtained as
 \begin{eqnarray}
\left[
\frac{1}{2kg_5^2}\left(1+\frac{\beta}{k^{3/2}}\widehat{\Phi}_{\text{OR}}(\pi) \right)+\frac{\beta_{\text{IR}}}{4g_{\text{IR}}^2k^{3/2}}\widehat{\Phi}_{\text{OR}} '(\pi)\right]\frac{\widetilde{\varphi}}{f}\:F_{\mu\nu} F^{\mu\nu} \; .
\label{eq:bulk-photon-withGW}
 \end{eqnarray} 
In this expression we have dropped the negligible contribution 
proportional to the $\alpha$ term in $\widehat{\Phi}$ that only receives support from the boundary region. This classical contribution must be added to the contribution arising from quantum effects, which remains of the same form as Eq.~\eqref{eq:zeromode-GB-rad-loop}.

From Eq.~\eqref{eq:rad-genmass-relation} we see that the final term in 
brackets in Eq.~\eqref{eq:bulk-photon-withGW} scales as $m_\varphi^2/\Lambda_{\text{IR}}^2$. However, the other correction term proportional to $\widehat{\Phi}_{\text{OR}}(\pi)$ does not appear to scale in a simple way with the radion mass. In order to understand the presence of this term, it is useful to consider the holographic interpretation of this scenario. In the dual description, sourcing the GW scalar on the UV brane corresponds to a deformation of the CFT by a primary operator. This deformation affects the RG evolution of the gauge coupling. To understand this in detail, we again need to relate the beta function for the gauge theory above the scale $\Lambda_{\rm IR}$, where the conformal symmetry is broken, to the parameters of the extra dimensional theory. When $\Phi$ acquires a VEV, the coupling of the GW field to the gauge bosons affects the gauge kinetic terms in the 5D construction, and hence the 4D gauge coupling in the dual theory. Since $\widehat{\Phi}$ depends on the location in the extra dimension, the beta function coefficient $b_>$ in 
the dual theory is affected, and now depends on the energy scale.

To determine the new $b_>$, we must once again obtain the correction to 
the brane localized gauge kinetic term as the location of the UV brane 
is moved. We separate the 5D integral over $\theta$ into two parts, one 
from $0$ to $\theta_0$ and another from $\theta_0$ to $\pi$. After 
integrating out the part of the extra dimension corresponding to $\theta 
< \theta_0$, we match to the theory with the lower cutoff. Then the 
gauge coupling at the scale $\Lambda_0$ corresponding to 
$\theta=\theta_0$ is given by
 \begin{eqnarray}
\frac{1}{g_{\text{UV}}^2(\Lambda_0)}=\frac{1}{g_{\text{UV}}^2}
+\frac{2 r_c}{g_5^2}
\int_0^{\theta_0}\:d\theta \left[1+\beta\frac{\widehat{\Phi}}{k^{3/2}}\right].
\label{eq:}
\end{eqnarray}
 The beta function in 4D dual theory is given by
 \begin{eqnarray}
\frac{b_>}{8\pi^2}\equiv \frac{d}{d\log\Lambda_0}\frac{1}{g_{\text{UV}}^2(\Lambda_0)}
=-\frac{2}{kg_5^2}\left(1+\frac{\beta}{k^{3/2}}\widehat{\Phi}(\theta_0)\right).
\label{eq:beta-more-GW-mass}
 \end{eqnarray}
 We notice that $b_>$ now depends on $\theta_0$, and hence on $\Lambda_0$. The form of the contribution from scales below $\Lambda_{\rm IR}$ remains unaffected by the addition of the GW scalar. Therefore, the form of the term proportional to $b_<$ is unchanged.

By taking the limit $\theta_0\rightarrow \pi-\frac{1}{kr_c}$ in 
Eq.~\eqref{eq:beta-more-GW-mass} we can, in the limit of large $kr_c$, neglect the effects of boundary region of $\widehat{\Phi}$ and obtain the value of $b_>$ just above the breaking scale. The full radion coupling is then obtained by combining this result with Eq.~\eqref{eq:bulk-photon-withGW} and Eq.~\eqref{eq:zeromode-GB-rad-loop} as
 \begin{eqnarray}
\left[
\frac{b_<-b_>}{32\pi^2}
+\frac{k^{-3/2}}{2}\widehat{\Phi}_{\text{OR}}'(\pi)\left( \frac{\beta}{kg_5^2} +\frac{\beta_{\text{IR}}}{2g_{\text{IR}}^2}\right)\right]
\frac{\widetilde{\varphi}}{f}\:F_{\mu\nu}^2 \;.
\label{eq:bulk-GB-2rad-GW-mass}
 \end{eqnarray} 
It follows from Eq.~\eqref{eq:bulk-photon-withGW} that, in general, the 
correction to the radion couplings from effects associated with 
stabilization of the extra dimension can be large. However, we see from 
Eq.~\eqref{eq:beta-more-GW-mass} that in the presence of the GW scalar, 
the identification of $b_>$ on the CFT side of the correspondence in terms of parameters on the AdS side is also modified. As can be seen from Eq.~\eqref{eq:bulk-GB-2rad-GW-mass}, when this effect is incorporated the correction to the form of the radion coupling is proportional to $\widehat{\Phi}_{\text{OR}}'(\pi)k^{-3/2}$. From Eq.~\eqref{eq:rad-genmass-relation} it follows that this scales as $m_\varphi^2/\Lambda_{\text{IR}}^2$, in agreement with results from the CFT side of the correspondence. It also follows from naive dimensional analysis (NDA) estimates of the sizes of the brane and bulk gauge couplings (see Appendix~\ref{app:c} for details) that the overall size of the correction is parametrically one-loop suppressed. This differs from the case of brane localized gauge bosons~\cite{Chacko:2013dra}, but agrees with the dual result for elementary gauge bosons in the 4D CFT~\cite{Chacko:2012sy}.

\section{Massive Gauge Bosons}
\label{sec:radion2gauge-bosons-massive}

In this section we determine the corrections to the radion couplings to the massive gauge bosons of the SM, the $W^{\pm}$ and the $Z$. As in the previous section, we consider SM gauge bosons residing in the bulk of the space and the SM Higgs field $H$ localized on the IR brane. For simplicity, we assume that Higgs-radion mixing is absent, which is natural if, for example, the Higgs is a pNGB. When the Higgs acquires a VEV, the $W^\pm$ and $Z$ gauge bosons become massive. The coupling of the radion to the field strength tensor squared can be determined just as in Sec.~\ref{sec:radion2gauge-bosons-massless}. In this case, however, because the gauge symmetry is broken, the radion can also have a nonderivative coupling to the gauge fields of the form $\varphi W_\mu W^\mu$. Since this is an operator of lower dimension than $\varphi F_{\mu \nu} F^{\mu \nu}$, it constitutes the dominant effect at low energies. In this section, we focus on couplings of this form.
 
We first determine the couplings in the absence of stabilization. The action, in addition to gauge kinetic terms of Eq.~\eqref{eq:gauge-kt-bulkbrane}, includes the brane localized operator
 \begin{eqnarray} 
\mathcal{S} &\supset& \int d^4xd\theta \delta(\theta-\pi) \sqrt{-G_{\text{IR}}}\: 
G_{\text{IR}}^{\mu\nu}(\mathcal{D}_\mu H)(\mathcal{D}_\nu H)^\dagger\: ,
\label{eq:bulk-gb-higgs-brane} 
 \end{eqnarray} 
where $\mathcal{D}_\mu=\partial_\mu-i W_\mu$ represents the gauge covariant derivative, and $W_\mu$ represents any massive gauge boson.  After replacing $H$ by its VEV, the operator in Eq.~\eqref{eq:bulk-gb-higgs-brane} generates a mass $m_W$ for the zero mode gauge bosons $W_\mu$. To zeroth order in $m_W^2/\Lambda_{\rm IR}^2$, the profile for the zero mode gauge boson is a 
constant~\cite{Csaki:2002gy,Hewett:2002fe}. Therefore, to obtain the
couplings of the zero mode $W_\mu(x)$, we can simply replace $W_\mu(x,\theta)$ by $W_\mu(x)$ in the action and integrate over the extra dimension.

To determine the coupling of the radion to the zero mode, we follow the 
same steps as in Sec.~\ref{sec:radion2gauge-bosons-massless}. The leading 
coupling in this case comes from the operator Eq.~\eqref{eq:bulk-gb-higgs-brane} itself and is given 
by~\cite{Rizzo:2002pq,Csaki:2007ns}
 \begin{eqnarray} 
\frac{2m_W^2}{g_4^2}\frac{\widetilde{\varphi}}{f}W_\mu W^\mu\:,
\label{eq:rad2W-leading} 
 \end{eqnarray} 
 where the index on $W$ is raised by $\eta_{\mu\nu}$. We see that this coupling has the same form as for the case when $W_\mu$ is localized on the visible brane.

In the presence of the GW scalar $\Phi$, there are additional operators in the action involving couplings between the gauge bosons and $\Phi$. In addition to operators of the form Eq.~\eqref{eq:GW-bulkGB-int} that lead to corrections to the 4D gauge coupling as in Eq.~\eqref{eq:gauge-coup-withGWscalar}, we consider the operator
 \begin{align}
\mathcal{L}_{\text{int}}= 
\beta_W\:\sqrt{-G_{\text{IR}}}\:\delta(\theta-\pi)\:G_{\text{IR}}^{\mu\nu} (\mathcal{D}_\mu H)(\mathcal{D}_\nu H)^\dagger\:\frac{\Phi}{k^{3/2}}\:
\label{betaWm}
\end{align} 
 where $\beta_W$ is a dimensionless number. When $\Phi$ gets a VEV, this term corrects the mass $m_W$ of the bulk gauge boson, which is now given by
 \begin{eqnarray}
m_W^2 = \widehat{m}_W^2\left(1+\beta_W\frac{\widehat{\Phi}(\pi)}{k^{3/2}}\right) \; .
 \end{eqnarray}
In this expression $\widehat{m}_W$ represents the gauge boson mass that arises from Eq.~\eqref{eq:bulk-gb-higgs-brane} in the absence of the correction term Eq.~\eqref{betaWm}. In the presence of this operator, the coupling of the radion also receives corrections taking the form
 \begin{eqnarray} 
\frac{m_W^2}{g_4^2}\frac{\widetilde{\varphi}}{f}W^2\left[2- \frac{\beta_W\widehat{\Phi}_{\text{OR}}'(\pi)} {k^{3/2}+\beta_W\widehat{\Phi}(\pi)}\right]\:. 
 \end{eqnarray} 
In this expression $m_W$ and $g_4$ are the corrected mass and gauge coupling. From Eq.~\eqref{eq:rad-genmass-relation} we see that the correction term scales as $m_\varphi^2/\Lambda_{\text{IR}}^2$, and is small if the radion is light.

\section{Bulk Fermions}
\label{sec:radion2fermions}

In this section we determine the couplings of the radion to SM fermions. 
For concreteness we focus on the interactions of the radion with the 
up-type quarks, the generalization to other SM fermions being 
straightforward. We consider a scenario where these fields emerge from 
bulk fermions $\mathcal{Q}$ and $\mathcal{U}$, and obtain their masses 
from a brane-localized Higgs $H$. As in the previous sections, we first 
obtain the radion couplings in the absence of a stabilization mechanism, 
and then we show how these results are modified in the presence of the 
GW field. We also obtain the holographic interpretation of the results.

\subsection*{Radion Couplings In the Absence of a Stabilization Mechanism}

In the absence of any dynamics that fixes the brane spacing, the 
relevant part of the action takes the form
 \begin{align}
\int\!\! d^4 x\!\! \int_0^\pi\!\!\!\! d\theta 
\Bigg[\!
\sqrt{G}\left(
\frac{i}{2}e_a^M
\overline{\mathcal{Q}}\Gamma^a\overleftrightarrow{\partial_M}\mathcal{Q}
-kc_q\overline{\mathcal{Q}}\mathcal{Q} + \mathcal{Q}\rightarrow\mathcal{U} \!
\right)\!+\!\sqrt{-G_{\text{IR}}}\delta(\theta-\pi) \left(
\frac{Y}{k}\overline{\mathcal{Q}}H\mathcal{U}+\text{h.c.}\!
\right)\!\!
\Bigg]\, ,
\label{eq:4.14}
 \end{align} 
 where $\overleftrightarrow{\partial} \equiv 
\overrightarrow{\partial}-\overleftarrow{\partial}$. The dimensionless 
parameters $Y$ and $c_q,\,c_u$ represent the brane localized Yukawa 
coupling and the bulk mass parameters for the 5D fermions respectively. 
For simplicity, we suppress all flavor indices. The $e_a^M$ represent 
the vielbein and $\Gamma^a$ the matrices that realize the 5D Clifford 
algebra.

In the absence of a VEV for the brane-localized Higgs, the boundary 
conditions on the 5D fermions $\mathcal{Q}$ and $\mathcal{U}$ are 
chosen such that each has a zero mode with the 
appropriate chirality. These zero modes are identified with the 
corresponding massless quarks in the SM before electroweak symmetry 
breaking. Once the Higgs gets a VEV, these modes acquire a mass. At the same time, a mixing is induced between the zero mode of 
$\mathcal{Q}$ and the KK modes of $\mathcal{U}$ and vice versa. As a 
result, the zero modes are not mass eigenstates. To work in a mass 
diagonal basis, a ``mixed'' KK decomposition can be 
performed~\cite{Azatov:2008vm}. In this basis the 5D fermions are 
expanded as
 \begin{align}
\mathcal{Q}(x,\theta) &= 
\begin{pmatrix}
	\mathcal{Q}_L(x,\theta) \\
	\mathcal{Q}_R(x,\theta)
\end{pmatrix}
=
\begin{pmatrix}
	Q_L^0(\theta) q_L^0 (x)+ Q_L^1(\theta) q_L^1 (x) +\ldots \\
	Q_R^0(\theta) u_R^0 (x)+ Q_R^1(\theta) q_R^1 (x) +\ldots
\end{pmatrix}
\nonumber \\
\mathcal{U}(x,\theta) &= 
\begin{pmatrix}
	\mathcal{U}_L(x,\theta) \\
	\mathcal{U}_R(x,\theta)
\end{pmatrix}
\,=
\begin{pmatrix}
	U_L^0(\theta) q_L^0 (x)+ U_L^1(\theta) u_L^1 (x) +\ldots \\
	U_R^0(\theta) u_R^0 (x)+ U_R^1(\theta) u_R^1 (x) +\ldots
\end{pmatrix}\:,
\label{eq:mxdkk}
\end{align}
 where the subscripts $L,R$ refer to 4D chiralities. In our notation the lower case letters represent the 4D fields, while the upper case letters represent their profiles in the bulk. The superscripts $0,1,\ldots$ refer to the mode number in the KK expansion. Using this expansion, the 4D fields in the spectrum are 
 \begin{eqnarray}
(q^0_L, u^0_R), (q^1_L, q^1_R), (u^1_L, u^1_R), \ldots \:.
\label{eq:fexp}
\end{eqnarray} 
Notice that the first term in the KK expansion of $\mathcal{Q}_R$ contains 
the 4D field $u_R$ which is where the mixed nature of KK 
decomposition manifests itself. The profiles $Q^i_{L,R},U^i_{L,R}$ can 
be solved for in this decomposition, and the details are given in 
Appendix~\ref{app:a}. The calculation of the profiles fixes the 
mass $m_f$ for the pair $(q^0_L, u^0_R)$ and the masses of the other KK 
modes in terms of the parameters of the 5D theory. We will take the 
KK scale $m_{\text{KK}} \sim k e^{-k\pi r_c}$ to be parametrically 
larger than the zero mode fermion masses $m_f$ and work to lowest 
order in $m_f/ m_{\text{KK}}$.

To obtain the coupling of the radion to the zero modes, we write the 
5D metric $G_{MN}$ and the vielbein $e_a^M$ in terms of $\varphi$ and 
expand about the VEV $\left<\varphi\right>=f$. Using the expressions for 
the profiles, the couplings of the radion can be determined as shown in 
Appendix \ref{app:a}. The final result takes the form
 \begin{eqnarray}
\mathcal{L} \supseteq -m_f\left(I_q+I_u\right)
\frac{\widetilde{\varphi}}{f}
\:\left(q_L^\dagger u_R + \text{h.c.}\right)\;,
\label{eq:zeromode-fermion-radion-noGW}
 \end{eqnarray}
 where $I_q$, $I_u$ are dimensionless numbers given by an overlap 
integral involving the profiles, and depend on the dimensionless 5D mass 
parameters $c_q$, $c_u$ respectively.\footnote{Throughout our analysis, 
we work in the regime where $c_q>-1/2$ and $c_u < 1/2$, because in the 
opposite regime, the behavior of the zero mode spectrum is qualitatively 
different~\cite{Contino:2004vy}.}  In what 
follows we will choose positive chirality for $\mathcal{Q}$ and negative 
chirality for $\mathcal{U}$, so the expression for $I_u$ may be 
obtained from that for $I_q$ by making the replacement $c_q\rightarrow -c_u$.
 To leading order in $e^{-k\pi r_c}$, the quantity $I_q$ is
 \begin{align}
I_q = \frac{1/2-c_q}{1-e^{-(1-2c_q)kr_c\pi}}
+c_q
\approx
\left\{
\begin{array}{ll}
c_q \;\; &,\;\;  c_q>\frac12
\\
&
\\
\frac12 \;\; &,\;\; c_q<\frac12 \;\;\;\;
\end{array}
\right.
\label{eq:Iq}
\end{align}
 where we have taken the two limits in which the expression simplifies 
considerably. Therefore, if $c_q<1/2$ and $c_u>-(1/2)$, the radion 
coupling scales as $-m_f(c_q-c_u)$. In the opposite regime, $c_q>1/2$ 
and $c_u<-(1/2)$, the coupling scales as $-m_f$. This agrees with the
existing results in the literature~\cite{Csaki:2007ns}.

How do we understand this result from the dual point of view? Recall 
that AdS/CFT relates the extra-dimensional coordinate $\theta$ to the RG 
scale $\mu$ in the dual theory. A 5D fermion $\Psi$ in AdS space 
corresponds to a fermionic CFT operator $\mathcal{O}_\Psi$. The value of 
the fermion field $\Psi$ at the boundary of AdS space, which we denote 
by $q_s(x)$, is identified with the source for the operator 
$\mathcal{O}_\Psi$. Therefore, the 4D CFT Lagrangian contains the term
 \begin{eqnarray}
\delta \mathcal{L} = \left.\Psi(x)\right|_{\text{AdS boundary}}\:\mathcal{O}_\Psi(x) 
\equiv q_s(x) \mathcal{O}_\Psi(x).
\label{eq:CFT-source}
 \end{eqnarray}
Because the 4D Dirac equation is first order, the boundary condition 
for the 5D field $\Psi$ must be subject to a chiral projection relating 
the left- and right-handed chiralities. Therefore only one of the two 
chiralities can be identified with the source. The 5D (dimensionless) 
mass parameter $c_\Psi$ is related to the scaling dimension 
$\Delta_{\Psi}$ of $\mathcal{O}_\Psi$ by
 \begin{eqnarray}
\Delta_{\Psi} = \left|c_\Psi\pm\frac12\right|+\frac32\:,
\label{eq:mass-op-dim-fermion}
 \end{eqnarray}
 where $\Psi = \mathcal{Q}, \, \mathcal{U}$ and the $\pm$ denotes the two 
choices for the chirality of the source~\cite{Contino:2004vy}.

The correspondence can be extended to the case of AdS space with two 
branes, thereby allowing a holographic interpretation of our results. 
The source $q_s(x)$ now becomes dynamical, being promoted to an 
elementary field that couples weakly to the CFT. Since, in general, the 
coupling to an elementary field constitutes an explicit breaking of the 
CFT, we expect that other conformal symmetry violating operators will be 
generated and will be present in the theory at an arbitrary 
renormalization scale $\mu$. These are represented by higher dimensional 
operators on the UV brane that are suppressed by powers of $k$. The 
operator in Eq.~\eqref{eq:CFT-source} generates a mixing between the CFT 
states and elementary field $q_s(x)$. The presence of the IR brane in 
AdS corresponds to the spontaneous breaking of the CFT, and leads to a 
mass gap in the spectrum. As a consequence of Eq.~\eqref{eq:CFT-source}, 
the mass eigenstates are mixtures of the elementary state $q_s(x)$ and 
composites that arise from the CFT dynamics.

In the mass diagonal basis prior to electroweak symmetry breaking, the 
spectrum in the 4D theory contains a massless chiral fermion 
corresponding to the zero mode of the 5D field. The localization of the 
zero mode in AdS space is governed by the mixing between $q_s$ and 
$\mathcal{O}_\mathcal{Q}$ in the dual picture. If the scaling dimension 
$\Delta_{\mathcal{Q}}$ is less than $5/2$, the operator in 
Eq.~\eqref{eq:CFT-source} is relevant and therefore large at low 
energies. As a result, the massless mode is mostly composite. Using 
Eq.~\eqref{eq:mass-op-dim-fermion}, this corresponds to the case of the 
corresponding 5D mass parameter being less than $1/2$ (we are focusing 
on the case of $\mathcal{Q}$ which has positive chirality) and results 
in the zero mode being localized toward the IR brane. Similarly, if 
the scaling dimension $\Delta_{\mathcal{Q}}$ is more than $5/2$, the 
operator in Eq.~\eqref{eq:CFT-source} becomes irrelevant, and, as a 
result, the mixing is small at low energies. Consequently, the massless 
mode is mostly elementary and corresponds to the 5D mass parameter 
$c_\Psi$ being greater than $1/2$ using Eq.~\eqref{eq:mass-op-dim-fermion}. This translates into the zero mode being localized toward the UV brane.

The form of the coupling of the dilaton to light fermions has been 
obtained~\cite{Chacko:2012sy,Bellazzini:2012vz}, both for the case when 
the fermions are mostly composite and the case when they are mostly 
elementary. In the first case, the dilaton coupling simply scales like 
$m_f$, which agrees with the result from Eq.~\eqref{eq:Iq}. In the other 
case, the coupling scales as 
$m_f(\Delta_{\mathcal{Q}}+\Delta_{\mathcal{U}}-4)$. This coefficient can 
be rewritten as
 \begin{eqnarray}
\Delta_{\mathcal{Q}}+\Delta_{\mathcal{U}}-4 = c_q-c_u = I_q + I_u \;\;\; {\rm for}\; c_q > 1/2, \; c_u < -1/2
\label{eq:c-delta}
 \end{eqnarray}
 where the first equality employs Eq.~\eqref{eq:mass-op-dim-fermion}, 
and the second Eq.~\eqref{eq:Iq}. The case of one composite and one 
elementary fermion is a straightforward generalization. From this 
analysis, we see that in each case the radion coupling agrees with the 
corresponding result for the dilaton in the literature.

\subsection*{Corrections Arising from Stabilization}

We now include the effects of stabilizing the extra dimension. In 
general, the fields $\mathcal{Q}$ and $\mathcal{U}$ will couple to the GW 
field $\Phi$ in the bulk, resulting in corrections to the radion 
couplings. To leading order in $\Phi$, the interactions of the bulk 
fermions with the GW scalar are of the form
 \begin{align}
\frac{\Phi}{k^{3/2}}&\!
\left[
\sqrt{G}\!\left(\!
\alpha_q
\frac{i}{2}e_a^M
\overline{\mathcal{Q}}\Gamma^a\overleftrightarrow{\partial_M}\mathcal{Q}
-\beta_q\:kc_q \overline{\mathcal{Q}}\mathcal{Q}
+
\mathcal{Q}\rightarrow\mathcal{U}\! \right)\!
+\!
\sqrt{-G_{\text{IR}}}\delta(\theta-\pi)\alpha_y
\left(
\frac{Y}{k}\overline{\mathcal{Q}}H\mathcal{U}+\text{h.c.}\!
\right)\!
\right]
\label{eq:bulk-fermions-GW-coupling}
\end{align} 
 where $\alpha_q$, $\alpha_u$, $\beta_q$, $\beta_u$ and $\alpha_y$ are 
dimensionless couplings whose natural sizes are estimated in 
Appendix~\ref{app:c}. To calculate the coupling of the radion to the 
zero modes in the presence of these terms, we follow the same steps as 
before. First, we replace $\Phi$ by its VEV and perform the mixed KK 
decomposition. This fixes the mass $m_f$ for the zero mode pair $(q_L^0, 
u_R^0)$ and the KK modes and determines the profiles in terms of the 
other theory parameters. Next, we consider fluctuations of $\Phi$ about 
its VEV associated with fluctuations of the background radion field. The 
operators in Eq.~\eqref{eq:bulk-fermions-GW-coupling} generate corrections 
to the radion coupling of Eq.~\eqref{eq:zeromode-fermion-radion-noGW}. The 
details of the calculation are in Appendix~\ref{app:a}. Including these 
effects, the coupling has the form
 \begin{eqnarray}
\mathcal{L} \supseteq -m_f\left(I_q+I_u+I_h\right)
\frac{\widetilde{\varphi}}{f}
\:\left(q_L^\dagger u_R + \text{h.c.}\right)\;.
\label{eq:zeromode-fermion-radion-GW}
 \end{eqnarray} 
 The quantities $I_q$ and $I_u$ again arise from overlap integrals 
involving the profiles and reduce to the results in Eq.~\eqref{eq:Iq} when $\widehat{\Phi}$ is set to zero. The quantity $I_h$ originates from the 
brane localized term involving both $\Phi$ and $H$ and also vanishes if the $\widehat{\Phi}$ is set to zero. It is given by
 \begin{equation}
I_h = -\frac{\widehat{\Phi}_{\text{OR}}'(\pi)}{2} \left[\frac{2\alpha_y}{k^{3/2}+\alpha_y\widehat{\Phi}(\pi)}- \frac{\alpha_q}{k^{3/2}+\alpha_q\widehat{\Phi}(\pi)}- \frac{\alpha_u}{k^{3/2}+\alpha_u\widehat{\Phi}(\pi)}\right] \equiv-\frac{\widehat{\Phi}_{\text{OR}}'(\pi)}{2k^{3/2}}X_h  \:
\label{eq:Ih-GW}
 \end{equation}
 where $X_h$ is expected to be of order one by NDA as shown in Appendix 
\ref{app:c}. 

We next compute $I_q$, and, as before, the expression for $I_u$ is obtained 
from $I_q$ by making the replacement $c_q\rightarrow -c_u, 
\;\alpha_q\rightarrow \alpha_u,\; \beta_q\rightarrow \beta_u$. It is 
important to take into account the fact that, in addition to inducing 
the direct coupling of the radion to the fermions, $\widehat{\Phi}$ 
also modifies the leading order fermion bulk profiles. In 
Appendix~\ref{app:a} we obtain the solution for $I_q$ taking all these 
effects into account. The result may be found in 
Eq.~\eqref{eq:Iq_exact}.

We focus our attention on the phenomenologically interesting cases where 
the fermion profiles are peaked toward either the UV or IR brane. These 
represent fermions that are either mostly elementary or mostly composite, 
and correspond to generalizations of the unstabilized analysis considered 
previously. We shall refer to these cases as being UV localized or IR 
localized respectively. We define the quantity
 \begin{equation}
\widetilde{c}_q\equiv c_q+c_q\frac{(\beta_q-\alpha_q)\widehat{\Phi}_{\text{OR}}(\pi)}{k^{3/2}+\alpha_q \widehat{\Phi}_{\text{OR}}(\pi)} \, .
 \label{eq:ctilde}
\end{equation}  
To leading order in $\frac{d\phantom{kr_c}}{d(kr_c\theta)}\widehat{\Phi}\equiv\widehat{\Phi} '$ and $e^{-k\pi r_c}$, $I_q$ is given by
  \begin{equation}
I_q=\left\{\begin{array}{lc}
\displaystyle \widetilde{c}_q & \, , \; \displaystyle \text{UV Localized}\\
&\\
\displaystyle \frac12+\frac{\widetilde{c}_q(\beta_q-\alpha_q) \widehat{\Phi}_{\text{OR}}'(\pi)k^{3/2}} {(1-2\widetilde{c}_q)(k^{3/2}+\alpha_q\widehat{\Phi}_{\text{OR}}(\pi)) (k^{3/2}+\beta_q\widehat{\Phi}_{\text{OR}}(\pi))} & \, , \; \displaystyle \text{IR Localized}
\end{array}\right. .
 \end{equation}

Now that the functions $I_q$, $I_u$ and $I_h$ have been determined, the 
fermion coupling to the radion can be determined from 
Eq.~\eqref{eq:zeromode-fermion-radion-GW}. In the case of IR localized 
profiles, the sum of the $I$ functions is given by
 \begin{align}
1-\frac{\widehat{\Phi}_{\text{OR}}'(\pi)}{k^{3/2}} &\left[ \frac{X_h}{2} -\frac{\widetilde{c}_q(\beta_q-\alpha_q)k^3} {(1-2\widetilde{c}_q)(k^{3/2}+\alpha_q\widehat{\Phi}_{\text{OR}}(\pi)) (k^{3/2}+\beta_q\widehat{\Phi}_{\text{OR}}(\pi))} \right.\nonumber\\
&\left.+\frac{\widetilde{c}_u(\beta_u-\alpha_u)k^3}{(1+2\widetilde{c}_u)(k^{3/2} +\alpha_u\widehat{\Phi}_{\text{OR}}(\pi)) (k^{3/2}+\beta_u\widehat{\Phi}_{\text{OR}}(\pi))}\right]\:.
\label{eq:full-coupling-fer-GW}
 \end{align} 
 We see that the corrections to the unstabilized result of 
Eqs.~\eqref{eq:zeromode-fermion-radion-noGW} and~\eqref{eq:Iq} scale as 
$\widehat{\Phi}_{\text{OR}}'(\pi)k^{-3/2}$. From Eq.~\eqref{eq:rad-genmass-relation}, it follows that the corrections to 
the leading result are proportional to $m_\varphi^2/\Lambda_{\text{IR}}^2$, in 
good agreement with the CFT side of the correspondence. 

We now turn to UV localized profiles. The sum of the $I$ functions in 
this case is given by
 \begin{equation} 
\left(\widetilde{c}_q-\widetilde{c}_u\right) - 
\frac{\widehat{\Phi}_{\text{OR}}'(\pi)}{2k^{3/2}} X_h \label{eq:full-coupling-fer-GW-UV}
 \end{equation} 
 This contains two distinct types of corrections to the unstabilized 
result. The term proportional to $\widehat{\Phi}_{\text{OR}}'(\pi)k^{-3/2}$ scales as $m_\varphi^2/\Lambda_{\text{IR}}^2$, in line with our expectations from holography. On the other hand, the difference between $c$ and 
$\widetilde{c}$ contains a correction term proportional to 
$\widehat{\Phi}_{\text{OR}}(\pi)$. This term is expected to be somewhat large, and does not scale in a simple way with the radion mass. To understand this 
result, we consider the holographic dual of this scenario. In the 
presence of operators such as Eq.~\eqref{eq:bulk-fermions-GW-coupling}, 
the relation between $c_\Psi$ and $\Delta_\Psi$ is modified from 
Eq.~\eqref{eq:mass-op-dim-fermion}. Specifically, the effective scaling 
dimension $\Delta_\Psi$ changes with the RG scale and the corrections to 
its value become large close to the breaking scale. This effect must be 
taken into account when relating $\Delta_\Psi$ at the breaking scale to 
$c_\Psi$. The details of the calculation are presented in 
Appendix~\ref{app:b} and follow the approach presented 
in~\cite{Contino:2004vy}. We find that to leading order in 
$\widehat{\Phi}_{\text{OR}}'$, Eq.~\eqref{eq:mass-op-dim-fermion} generalizes to
 \begin{align}
\Delta_{\Psi} &= \left|c_\Psi\left[1+\frac{\left(\beta_\Psi-\alpha_\Psi\right) \widehat{\Phi}_{\text{OR}}(\pi)}{k^{3/2}+\alpha_q\widehat{\Phi}_{\text{OR}}(\pi)} \right]\pm\frac12\right|+\frac32+\mathcal{O}\left(k^{-3/2}\widehat{\Phi}_{\text{OR}}' \right)\nonumber\\
&= \left| \widetilde{c}_\Psi \pm\frac12 \right| +\frac32 + \mathcal{O}\left(k^{-3/2}\widehat{\Phi}_{\text{OR}}' \right)
\label{eq:hologmass}
 \end{align}
 where $\Psi=\mathcal{Q},\,\mathcal{U}$ and the operator dimension 
$\Delta_{\Psi}$ in this expression is understood to be evaluated close 
to the symmetry breaking scale. As before, the $\pm$ denotes the two 
choices of chirality, and we choose it to be positive for $\mathcal{Q}$ 
and negative for $\mathcal{U}$.

Using this modified relation and neglecting terms of order $k^{-3/2}\widehat{\Phi}_{\text{OR}}'$, we find that 
 \begin{equation}
c_q\left[1+\frac{\left(\beta_q-\alpha_q\right) \widehat{\Phi}_{\text{OR}}(\pi)}{k^{3/2}+\alpha_q\widehat{\Phi}_{\text{OR}}(\pi)} \right]-c_u\left[1+\frac{\left(\beta_u-\alpha_u\right) \widehat{\Phi}_{\text{OR}}(\pi)}{k^{3/2}+\alpha_u\widehat{\Phi}_{\text{OR}}(\pi)} \right]
=\widetilde{c}_q-\widetilde{c}_u= \Delta_\mathcal{Q}+\Delta_{\mathcal{U}}-4\:.
\label{eq:mod-c2delta-GW}
\end{equation}
 As a result, the large term that scales as $\widehat{\Phi}_{\text{OR}}(\pi) $ for UV profiles in Eq.~\eqref{eq:full-coupling-fer-GW-UV} is absorbed into 
$\Delta_\mathcal{Q}$ and $\Delta_\mathcal{U}$ when the dilaton coupling is 
written in terms of the operator dimensions at the breaking scale. The 
remaining corrections scale as $\widehat{\Phi}_{\text{OR}}'(\pi)k^{-3/2} \sim 
m_\varphi^2/\Lambda_{\rm IR}^2$, as expected from the CFT side of the 
correspondence~\cite{Chacko:2012sy}.

In summary, we see that the leading order radion couplings to bulk SM 
fermions correspond to dilaton interactions that scale either as $m_f$ 
or as $m_f\left(\Delta_\mathcal{Q}+\Delta_\mathcal{U}-4\right)$, 
depending on whether the SM fermions are mostly composite or mostly 
elementary. In the presence of the GW field, the identification of 
$\Delta_\mathcal{Q}$ and $\Delta_\mathcal{U}$ with parameters in the 
dual 5D theory receives corrections. When this effect is taken into 
account, the leading corrections to the form of the dilaton interaction 
are found to scale as $m_\varphi^2/\Lambda_{\text{IR}}^2$, in good agreement with results from the CFT side of the correspondence.

\section{Conclusion}
\label{sec:conclusion}

%We have calculated the coupling of the radion to the zero modes of bulk 
%fields in the presence of stabilization. We find that for every field 
%the corrections to the couplings, relative to the unstabilized analysis, 
%scale as the squared ratio of the mass of the radion to KK scale.

The AdS/CFT correspondence is a powerful tool that can help us 
understand the dynamics of strongly coupled 4D theories by studying 
their weakly coupled higher dimensional duals. A particularly 
interesting laboratory to study the duality is in the context of the 
explicit and spontaneous breaking of the isometries of the extra 
dimensions, corresponding to the spontaneous breaking of an approximate 
conformal symmetry in the 4D theory. The spontaneous breaking gives 
rise to an associated Goldstone boson, the radion in the extra 
dimension and dilaton in the CFT. In this work we have studied the 
interactions of a radion in a class of theories of phenomenological 
interest, specifically RS models with the SM gauge and matter fields in 
the bulk. We have compared the results against those in the literature 
for the dilaton, finding good agreement.

In the absence of a stabilization mechanism for the extra dimension 
such as the GW framework~\cite{Goldberger:1999uk}, the form of the 
radion couplings is determined by diffeomorphism invariance. Here, we 
have computed the corrections to these couplings that arise from the 
stabilization mechanism. We have focused on the phenomenologically 
interesting case where the radion is somewhat lighter than the KK 
states associated with the extra dimension. We have extended the 
analysis of~\cite{Chacko:2013dra}, which was restricted to the scenario 
when all the SM fields were localized to the IR brane, to the case when 
the gauge bosons and fermions of the SM reside in the bulk of the extra 
dimension. These corrections primarily arise from direct couplings of 
the GW scalar to the SM fields of the form shown schematically in 
Eq.~\eqref{eq:bulk-op-phi-in-bulk}.

We have obtained a detailed interpretation of our results in terms of 
the holographic dual of the radion, the dilaton. In doing so, we have 
taken into account the fact that the familiar identification of the 
parameters on the two sides of the AdS/CFT correspondence is modified 
in the presence of couplings of the bulk SM fields to the GW scalar. As 
in the case of brane-localized SM fields, we find that corrections to 
the form of the dilaton couplings to these states are suppressed by the 
square of the ratio of the dilaton mass to the KK scale. 
These effects are therefore parametrically small in the limit of a 
light radion, in good agreement with the corresponding results for the 
dilaton~\cite{Chacko:2012sy}.

\acknowledgments
Z.C., R.K.M. and C.B.V. are supported by the NSF under Grant No. PHY-1315155. 
%\newpage
\appendix

\section{Radion Mixing with the GW Field}
\label{app:mixing}

In general, the GW stabilization mechanism will affect the radion profile, leading to corrections to its couplings. In the KK picture, these changes in the radion wave function arise from mixing between the radion and other states after stabilization. In the limit of small backreaction, the leading corrections to the radion profile are expected to arise from mixing with the KK modes of the GW scalar, rather than from mixing with the graviton or its KK modes. The physical radion state is then a linear combination of the graviscalar and these heavy scalar fields. Consequently, the radion couplings to the SM fields receive corrections. In this appendix we determine the size of these effects. In particular, we show that they are smaller than the corrections that arise from direct couplings of the GW scalar to SM fields.

We begin from the action for the GW scalar
 \begin{align}
\mathcal{S}_{\text{GW}}
&=\int d^4xd\theta
\left[
\sqrt{G}\left(
\frac12G^{AB}\partial_A\Phi\partial_B\Phi
-V_b(\Phi)
\right)\!
-\!\!\sum_{i=\text{IR,UV}}\delta(\theta-\theta_i)\sqrt{-G_i}V_i(\Phi)
\right].
\label{eq:Appgw-action}
 \end{align}
 We now make a change of variables from $\Phi(x,\theta)$ to a new 
variable $\phi(x,\theta)$, by making the separation $\Phi(x,\theta) = 
 \widehat{\Phi}(r(x),\theta)  + \phi(x,\theta)$. Here $\widehat{
\Phi}(r(x),\theta)$ corresponds to the VEV of $\Phi$ at the 
minimum, but with $r_c$ promoted to the dynamical field $r(x)$. Having made this change of variables, we substitute for $\Phi(x,\theta)$ in the action. Because $\widehat{\Phi}$ satisfies the classical equations of motion, several terms in the action cancel. We are left with
 \begin{align}
\mathcal{S}_{\text{GW}}
&=\int d^4xd\theta \left[-\frac{e^{-4kr\theta}}{2r}\partial_{\theta} \phi\partial_{\theta}\phi +\frac{re^{-2kr\theta}}{2}\left(\partial_{\mu}\widehat{\Phi} \partial^{\mu}\widehat{\Phi}+ 2\partial_{\mu}\widehat{\Phi} \partial^{\mu}\phi+ \partial_{\mu}\phi \partial^{\mu}\phi \right)\right.\nonumber\\
&-re^{-4kr\theta}\left( \frac12\phi^2\frac{\partial^2} {\partial\widehat{\Phi}^2}V_b(\widehat{\Phi}) +\ldots\right) -\delta(\theta)\left( \frac12\phi^2\frac{\partial^2} {\partial\widehat{\Phi}^2}V_{\text{UV}}(\widehat{\Phi})+\ldots\right) \nonumber\\
 &\left.-\delta(\theta-\pi)e^{-4kr\pi}\left( \frac12\phi^2\frac{\partial^2} {\partial\widehat{\Phi}^2}V_{\text{IR}}(\widehat{\Phi})+\ldots\right)\right]  \; ,
 \end{align}
 where the $+\ldots$ represent terms higher order in $\phi$. We neglect 
these higher order terms, since their effects are subleading. In 
addition, we replace $r(x)$ with $r_c$ in terms that are quadratic in 
$\phi$, or that involve $\partial_{\mu}\widehat{\Phi}\sim\partial_{\mu}r(x)$, 
since the effects being neglected are small. After these 
simplifications, the relevant part of the action takes the form
 \begin{align}
\mathcal{S}_{\text{GW}}
&=\int d^4xd\theta \left[-\frac{e^{-4kr_c\theta}}{2r_c}\partial_{\theta} \phi\partial_{\theta}\phi+\frac{r_ce^{-2kr_c\theta}}{2} \left(\partial_{\mu}\widehat{\Phi} \partial^{\mu}\widehat{\Phi}+ 2\partial_{\mu}\widehat{\Phi}\partial^{\mu}\phi+ \partial_{\mu}\phi \partial^{\mu}\phi \right)\right.\nonumber\\
&\left.-\frac{r_c}{2}e^{-4kr_c\theta}\phi^2\frac{\partial^2}{\partial\widehat{\Phi}^2} V_b(\widehat{\Phi}) -\delta(\theta) \frac12\phi^2\frac{\partial^2}{\partial\widehat{\Phi}^2}V_{\text{UV}}(\widehat{\Phi}) -\delta(\theta-\pi)e^{-4kr_c\pi}\frac12\phi^2\frac{\partial^2} {\partial\widehat{\Phi}^2}V_{\text{IR}}(\widehat{\Phi})\right]  
 \end{align}
 We see from the form of the action that the only mixing between the 
light graviscalar and the heavy modes contained in $\phi$ arises from 
the kinetic terms. To determine the size of this effect we employ the KK 
decomposition, $\phi(x,\theta)=\sum f_n(\theta)\phi_n(x)$ in the classical 
background, $r=r_c$. The profiles $f_n$ satisfy the equation
 \begin{equation}
\partial_{\theta}\left(e^{-4kr_c\theta}\partial_{\theta}f_n \right) -\frac{r_c^2}{2}e^{-4kr_c\theta}\frac{\partial^2V_b}{\partial \widehat{\Phi}^2}f_n =-m_n r_c^2e^{-2kr_c\theta}f_n,
 \end{equation}
subject to the boundary conditions 
 \begin{align}
\partial_{\theta}f_n&=r_cf_n\frac{\partial^2V_{\text{UV}}}{\partial \widehat{\Phi}^2} &\theta=0,\nonumber\\
-\partial_{\theta}f_n&=r_cf_n\frac{\partial^2V_{\text{IR}}}{\partial \widehat{\Phi}^2} &\theta=\pi \;.
 \end{align} 
 It is convenient to normalize these profiles as
 \begin{equation}
\int d\theta r_ce^{-2kr_c\theta}f_nf_m=\delta_{nm}.
 \end{equation}
 Then the action reduces to
 \begin{align}
\mathcal{S}_{\text{GW}}
&=\int d^4x\left[\sum_n\left(\frac12 \partial_{\mu}\phi_n\partial^{\mu}\phi_n -\frac{m_n^2}{2}\phi_n^2 \right)\nonumber\right.\\
&+\left.\int d\theta\left(\frac{r_c}{2}e^{-2kr_c\theta}\partial_{\mu}\widehat{\Phi} \partial^{\mu}\widehat{\Phi} +\sum_n r_ce^{-2kr_c\theta}f_n\partial_{\mu}\widehat{\Phi}\partial^{\mu} \phi_n \right)\right].\label{eq:kineticcorrections}
 \end{align}

At this point we recall that the $x$ dependence of $\widehat{\Phi}$ arises through
$r(x)$,
 \begin{equation}
\partial_{\mu}\widehat{\Phi}=\frac{\partial\widehat{\Phi}}{\partial r}\partial_{\mu}r=-\frac{1}{k\pi \varphi}\frac{\partial\widehat{\Phi}}{\partial r}\partial_{\mu}\varphi,
 \end{equation} 
 where we have made the change of variable from $r(x)$ to the 
canonically normalized radion field $\varphi=\sqrt{24M_5^3/k}\exp(-k\pi r(x))$. 
We see that Eq.~\eqref{eq:kineticcorrections} contains, in addition to a correction to the radion kinetic term, a term that generates kinetic mixing between the radion and the KK states of the GW field. Now, from \eqref{eq:phi-soln-massANDcubic} we have
 \begin{equation}
\frac{\partial\widehat{\Phi}}{\partial r}=k^{5/2}\alpha (\pi -\theta)e^{-4kr(\pi-\theta)} +k\theta\widehat{\Phi}_{\text{OR}}'.
 \end{equation} 
 Under the $\theta$ integrals the $\alpha$ term is exponentially 
suppressed except in the region close to $\theta=\pi$, where its 
coefficient is small. In what follows we neglect this term, since its contribution is small.

The coefficient of the correction to the radion kinetic term is given by
 \begin{equation}
 \frac{r_ck^3}{\pi^2\langle\varphi\rangle^2}\int d\theta e^{-2kr_c\theta}\theta^2 k^{-3}\widehat{\Phi}_{\text{OR}}'^2.\label{eq:radkincorrect}
 \end{equation} 
 Now, the VEV of $\Phi$ grows from the UV to the IR, and, in general, so 
does $\widehat{\Phi}_{\text{OR}}'$. Therefore, we expect that at some arbitrary 
point $\theta$ in the bulk, we have that $\widehat{\Phi}_{\text{OR}}'(k r_c 
\theta) \lesssim \widehat{\Phi}_{\text{OR}}'(k r_c \pi)$. This allows us to bound 
\eqref{eq:radkincorrect} as
 \begin{equation}
\frac{r_ck^3}{\pi^2\langle\varphi\rangle^2}
\int d\theta e^{-2kr_c\theta}\theta^2 k^{-3}\widehat{\Phi}_{\text{OR}}'^2
\lesssim
\frac{r_ck^3}{\pi^2\langle\varphi\rangle^2} k^{-3} \widehat{\Phi}_{\text{OR}}'(kr_c\pi)^2
\int d\theta e^{-2kr_c\theta}\theta^2.
 \end{equation}
 Up to exponentially suppressed terms the $\theta$ integral evaluates to 
$(2kr_c)^{-2}$. Noting that $k^{-3/2}\widehat{\Phi}_{\text{OR}}'(kr_c\pi) \sim 
m_{\varphi}^2/ \Lambda_{\text{IR}}^2$, we see that the correction to the 
radion kinetic term satisfies
 \begin{equation}
\frac{r_ck^3}{\pi^2\langle\varphi\rangle^2}\int d\theta e^{-2kr_c\theta}\theta^2 k^{-3}\widehat{\Phi}_{\text{OR}}'^2 \lesssim \left(\frac{1}{2\pi\langle\varphi\rangle r_c}\right) ^2\frac{m_{\varphi}^4}{\Lambda_{\text{IR}}^4} \; .
 \end{equation} 
 Since this correction scales as $m_{\varphi}^4/\Lambda_{\text{IR}}^4$, 
we see that its effect on the radion interactions is smaller than the corrections that arise from direct couplings of the GW field to the SM, which scale as $m_{\varphi}^2/\Lambda_{\text{IR}}^2$.

The mixing term takes the form
 \begin{equation}
-\int d^4x\frac{r_ck^{3/2}}{\pi \langle\varphi\rangle} \sum_n f_n\left[\int d\theta e^{-2kr_c\theta}\theta k^{-3/2}\widehat{\Phi}_{\text{OR}}' \right] \partial_{\mu} \varphi \partial^{\mu}\phi_n\equiv\int d^4x\sum_n\kappa_n\partial_{\mu} \varphi \partial^{\mu}\phi_n\, .
 \end{equation}
 Employing the same methods as in the previous case, we find that the coefficients $\kappa_n$ of the mixing terms satisfy 
$\kappa_n \lesssim m_{\varphi}^2/\Lambda_{\text{IR}}^2$. Upon transforming 
to a basis where the kinetic terms are diagonal and canonically 
normalized, we find that
 \begin{equation}
\begin{array}{lcr}
\displaystyle\varphi\to\varphi- \kappa_n\frac{m_n^2}{m_n^2-m_{\varphi}^2}\phi_n ,
& \phantom{\phi\to\phi}&
\displaystyle \phi_n\to \phi_n+\kappa_n\frac{m_{\varphi}^2}{m_n^2-m_{\varphi}^2} \varphi
\end{array}.
 \end{equation}
 The mass of the KK states of $\phi$ is of the order of the IR scale, 
$m_n\sim\Lambda_{\text{IR}}$. Then it follows that the corrections to 
the radion couplings that arise from mixing with the KK states of the GW 
scalar scale as $m_{\varphi}^4/\Lambda_{\text{IR}}^4$, and are, therefore, smaller than the effects from direct couplings to the GW field.

\section{Couplings of Bulk Fermions to the Radion}
\label{app:a}

In this appendix we determine the form of the radion coupling to bulk 
fermions in the presence of the GW field, filling in many of the steps 
outlined in Sec.~\ref{sec:radion2fermions}. Consistent with the 
metric Eq.~\eqref{eq:RS-metric}, we define the vielbein
 \begin{equation}
e_{a}^{M}
=
\delta_{a}^{\mu}\delta_\mu^M e^{kr_c|\theta|}+\frac{1}{r_c}\delta^5_{a}\delta_5^M \; .
\label{eq:veilbein}
 \end{equation}
 Here $M$ is the 5D curved index and $a$ is the 5D index in the 
tangent space. We choose the gamma matrices to be
 \begin{equation}
\begin{array}{ccc}
\gamma^{a}=(\gamma^{\mu},-i\gamma^5), &\displaystyle\gamma^{\mu}=\left(\begin{array}{cc}
0&\sigma^{\mu}\\
\bar{\sigma}^{\mu}&0
\end{array}\right), &\displaystyle\gamma^5=\left(\begin{array}{cc}
-\mathbb{I}&0\\
0&\mathbb{I}
\end{array}\right).
\end{array}
\label{eq:5dgamma}
\end{equation}
For later convenience, we choose to write a 5D fermion $\Psi$ as
\begin{equation}
\begin{array}{cc}
\displaystyle\Psi=\left(\begin{array}{c}
\Psi_L\\
\Psi_R
\end{array}\right),
\end{array}
 \end{equation}
 which fixes the form of $\overline{\Psi}= \Psi^\dagger\gamma^0 = 
\left(\Psi^\dag_R \; \Psi^\dag_L \right)$. For concreteness we focus on 
radion couplings to the up-type quarks. The generalization to the cases 
of the other SM fermions is straightforward.

In the presence of the GW scalar $\Phi$, the 5D fermion action is given by
\begin{align}
S=\!\int\! d^4x\!\int_0^\pi\!\! d\theta\sqrt{G}
&\left[
\frac{i}{2}\left(\overline{\mathcal{Q}}\Gamma^{M} \partial_{M}\mathcal{Q} 
-\left(\partial_{M}\overline{\mathcal{Q}} \right)\Gamma^{M} \mathcal{Q}\right)
\left(1+\frac{\alpha_q}{k^{3/2}}\Phi \right)
-m_{\mathcal{Q}}\overline{\mathcal{Q}}\mathcal{Q} 
\left(1+\frac{\beta_q}{k^{3/2}}\Phi \right)
\right.
\nonumber\\
&+\frac{i}{2}\left(\overline{\mathcal{U}}\Gamma^{M} \partial_{M}\mathcal{U} 
-\left(\partial_{M}\overline{\mathcal{U}} \right)\Gamma^{M} \mathcal{U}\right)
\left(1+\frac{\alpha_u}{k^{3/2}}\Phi \right)
- m_{\mathcal{U}}\overline{\mathcal{U}}\mathcal{U}
\left(1+\frac{\beta_u}{k^{3/2}}\Phi \right)
\nonumber\\
&\left.
+\frac{\delta(\theta-\pi)}{r_c}
\left(\frac{Y}{k}\overline{\mathcal{Q}}H\mathcal{U} +\text{h.c.} \right) 
\left(1+ \frac{\alpha_y}{k^{3/2}}\Phi 
\right)\right],
\end{align}
where $\Gamma^M=e_a^M\gamma^a$. For simplicity we take $Y$ to be real and consider the Higgs field $H$ to be localized to the visible brane at $\theta=\pi$. We denote the VEV of $H$ by $v_h$. 

The dynamics in the $\Phi$ sector leads to a background value 
$\widehat{\Phi}(\theta)$. The excitations of the GW field are generically heavy, 
being of order $m_{\text{KK}}$. This allows us to integrate out the GW field in 
a dynamical radion background, thereby obtaining the low energy 
effective theory for the radion. We do this by promoting $r_c$ to a 
dynamical field, which we denote by $r(x)$, and expanding $r(x)$ about $r_c$ as 
$r(x)=r_c+\delta r(x)$. Using Eq.~\eqref{eq:rad2r}, the canonically normalized physical radion $\widetilde{\varphi}$ is related to the other parameters by
\begin{equation}
\varphi=\langle\varphi\rangle+\widetilde{\varphi}=\sqrt{\frac{24M^3_5}{k}} e^{-kr_c\pi}(1-k\pi\delta r)
\end{equation}
which leads to the relation
\begin{equation}
\delta r=-\frac{\widetilde{\varphi}}{fk\pi}
\label{eq:rad2rfluct}
\end{equation}
where $\langle\varphi\rangle=f$.

To proceed, we take the following approach. We first set $r=r_c$ and 
determine the equations of motion for the 5D fermions. We then set the 
Higgs field to its VEV and perform a KK decomposition to obtain the 
4D fermion spectrum. We then write out the action to linear order in 
$\delta r$, expand out the bulk fermions in terms of their KK 
modes, and integrate over the extra dimension to obtain the radion 
coupling to the zero modes. This approach is consistent within the 
effective theory and yields the leading contribution to the operators 
that couple a single radion to the SM fermions.

To be consistent with phenomenology, we choose $\mathcal{Q}_L$ and 
$\mathcal{U}_R$ to be even about $\theta=0$ and $\mathcal{Q}_R$ and 
$\mathcal{U}_L$ to be odd. As usual, the mass parameter 
$c_i$ is taken to be odd. We also take $\widehat{\Phi}_c$, the VEV of $\Phi$ at $r=r_c$, to be even at $\theta=0$ and $\theta = \pi$. Using the orbifold symmetry, we restrict the limits on the $\theta$ integral in the action to be from $0$ to $\pi$.

Minimizing the action (at $r=r_c$), we find the equations of motion 
satisfied by the fermion fields. For instance, considering the variation 
$\delta \mathcal{Q}_R^{\dag}$ gives 
\begin{align} & 
ir_ce^{kr_c\theta}\sigma^{\mu}\partial_{\mu}\mathcal{Q}_R-\partial_\theta 
\mathcal{Q}_L +2kr_c\mathcal{Q}_L-kr_cc_q\frac{1+(\beta_q/k^{3/2}) \widehat{\Phi}_c}{1+(\alpha_q/k^{3/2})\widehat{\Phi}_c 
}\mathcal{Q}_L\nonumber\\ 
&\phantom{AAAAAAA} -\frac{(\alpha_q/2k^{3/2})\partial_{\theta}\widehat{\Phi}_c} {1+(\alpha_q/k^{3/2})\widehat{\Phi}_c}\mathcal{Q}_L 
+\frac{v_hY}{2k} 
\delta(\theta-\pi)\frac{1+(\alpha_y/k^{3/2})\widehat{\Phi}_c} {1+(\alpha_q/k^{3/2})\widehat{\Phi}_c} 
\mathcal{U}_L =0\: .
\label{eq:QLeq} 
\end{align} 
The boundary terms in the action fix the boundary conditions to be 
\begin{eqnarray} &&\left[\delta 
\mathcal{Q}_R^{\dag}\mathcal{Q}_Le^{-4kr_c\theta} 
\left(1+\frac{\alpha_q}{k^{3/2}}\widehat{\Phi}_c \right) +\frac{Yv_h}{2r_ck} 
\delta \mathcal{Q}_R^{\dag}\mathcal{U}_L e^{-4kr_c\theta} 
\left(1+\frac{\alpha_Y}{k^{3/2}}\widehat{\Phi}_c\right)\right]\Big|_{\theta=\pi} 
\nonumber \\ \qquad &&-\left[\delta 
\mathcal{Q}_R^{\dag}\mathcal{Q}_Le^{-4kr_c\theta} 
\left(1+\frac{\alpha_q}{k^{3/2}}\widehat{\Phi}_c \right)\right]\Big|_{\theta=0} 
=0 \, . 
\label{eq:QLbc} 
\end{eqnarray} 
We proceed by employing the mixed KK decomposition described in 
Eq.~\eqref{eq:mxdkk} and require the zero modes $q_L^0$ and $u_R^0$ to 
satisfy the 4D Dirac equations
\begin{equation} \begin{array}{cc} \displaystyle 
i\overline{\sigma}^{\mu}\partial_{\mu}q^0_L-m_fu^0_L=0\;, \;\;\; & \displaystyle 
i\sigma^{\mu}\partial_{\mu}u^0_R-m_f q^0_L=0\:, \end{array} 
\label{eq:mixedKK-zeromode} \end{equation} 
where $m_f$ is the mass of the zero mode generated by the Higgs VEV. In what follows, we work to leading order in $m_f/(ke^{-k\pi r_c})$. Because we have chosen $\mathcal{Q}_R$ and $\mathcal{U}_L$ to be odd about $\theta=0$, they 
vanish at the boundary. This ensures that the boundary condition at 
$\theta=0$ is satisfied in Eq.~\eqref{eq:QLbc}. To leading order in 
$m_f/(ke^{-k\pi r_c})$, this completely fixes the profiles $Q_L^0$ and 
$U^0_R$ up to an overall normalization that is determined by the requirement 
of a canonical kinetic term for the 4D field. The boundary condition 
at $\theta=\pi$, to this order, fixes the mass $m_f$ in terms of other 
parameters.

Using Eqs.~\eqref{eq:mixedKK-zeromode} and~\eqref{eq:QLeq}, the profile $Q_L^0$ satisfies
\begin{align}
\partial_\theta Q^0_L-\!
&\left(\!2kr_c-kr_cc_q\frac{\displaystyle 1+(\beta_q/k^{3/2})\widehat{\Phi}_c}{\displaystyle 1+(\alpha_q/k^{3/2})\widehat{\Phi}_c} -\frac{\displaystyle (\alpha_q/2k^{3/2}) \partial_{\theta}\widehat{\Phi}_c}
{\displaystyle 1+(\alpha_q/k^{3/2})\widehat{\Phi}_c}\!\right)\!Q^0_L \!
-m_fr_ce^{kr_c\theta}Q^0_R=0.
\label{eq:Q0L-profile-eqn}
 \end{align}
Similar equations can be derived for the other three fermion zero mode 
profiles $Q^0_R, \, U^0_{L,R}$. By our choice of boundary conditions, the 
even profiles $Q^0_L$ and $U^0_R$ correspond to the chiral fermions in 
the effective theory and hence survive in the $m_f\rightarrow 0$ limit. 
The odd profiles $Q^0_R$ and $U^0_L$ vanish at $\theta=0$ and are 
forced by the equations of motion to begin at order 
$m_f/(ke^{-kr_c\pi})$. As a result, we can drop the term proportional to 
$Q_R^0$ in Eq.~\eqref{eq:Q0L-profile-eqn}. To make the notation simpler, we 
define the functions
 \begin{eqnarray}
T(\alpha,\theta)&=&\int^{\theta}_0 d\theta '\frac{\widehat{\Phi}_c(\theta')} {k^{3/2}+\alpha\widehat{\Phi}_c(\theta ')},
\nonumber \\
G(c,\alpha,\beta,\theta)&=&\exp\left[kr_c\theta\left(\frac12-c\right) -kr_cc(\beta-\alpha)T(\alpha,\theta)\right]\:.
\label{eq:T}
\end{eqnarray}
The even profiles, to leading order in $m_f/(ke^{-kr_c\pi})$, are given by
 \begin{align}
Q^0_L
=&
\frac{N_{Q_L}}
{\displaystyle\sqrt{1+(\alpha_q/k^{3/2})\widehat{\Phi}_c}} 
\exp\left(\frac32 kr_c\theta\right)
G(c_q,\alpha_q,\beta_q,\theta)\label{eq:Qprofile}
\\
U^0_R
=&
\frac{N_{U_R}}
{\displaystyle\sqrt{1+(\alpha_u/k^{3/2})\widehat{\Phi}_c}} 
\exp\left(\frac32 kr_c\theta\right)
G(-c_u,\alpha_u,\beta_u,\theta) \, .
 \end{align}
In the limit where $\widehat{\Phi}_c$ goes to zero, these agree with the results for the profiles in the absence of stabilization~\cite{Grossman:1999ra,Gherghetta:2000qt}. 
The constants $N_{Q_L}$ and $N_{U_R}$ are determined by normalizing the 
kinetic terms for $q_L^0$ and $u_R^0$ and are given by
 \begin{align}
N_{Q_L}^{-2}
&=
2r_c\int_0^{\pi}d\theta\, G^2(c_q,\alpha_q,\beta_q,\theta) ,
\nonumber \\
N_{U_R}^{-2}
&=
2r_c\int_0^{\pi}d\theta\, G^2(-c_u,\alpha_u,\beta_u,\theta) 
\label{eq:Qnorm}\, .
\end{align}

To leading order in $m_f/(ke^{-kr_c\pi})$ the odd profiles are given by
 \begin{align}
Q^0_R(\theta)
&=
-\frac{m_fr_cN_{Q_L}}{\sqrt{\displaystyle 1+(\alpha_q/k^{3/2})\widehat{\Phi}_c}}
\exp\left(\frac32 kr_c\theta\right)G(-c_q,\alpha_q,\beta_q,\theta)
\int_0^\theta d\theta' G^2(c_q,\alpha_q,\beta_q,\theta '),
\nonumber\\
\nonumber \\
U^0_L(\theta)
&=
\frac{m_fr_cN_{U_R}}{\sqrt{\displaystyle 1+(\alpha_u/k^{3/2})\widehat{\Phi}_c}}
\exp\left(\frac32 kr_c\theta\right)G(c_u,\alpha_u,\beta_u,\theta)
\int_0^\theta d\theta' G^2(-c_u,\alpha_u,\beta_u,\theta ').
 \end{align} 
 The brane localized Higgs term contributes to the boundary condition at 
$\theta=\pi$. Since the Yukawa operator is associated with effects 
suppressed by $m_f/(ke^{-kr_c\pi})$, to the order we are working this 
only affects the odd profiles. More specifically, the boundary 
conditions in Eq.~\eqref{eq:QLbc} require
 \begin{align}
Q^0_R(\pi)&=\frac{v_hY}{2k}U^0_R(\pi) \left(1+\frac{\alpha_y}{k^{3/2}}\widehat{\Phi}_c(\pi) \right) \left(1+\frac{\alpha_q}{k^{3/2}}\widehat{\Phi}_c(\pi) \right)^{-1}, \\
U^0_L(\pi)&=-\frac{v_hY}{2k}Q^0_L(\pi) \left(1+\frac{\alpha_y}{k^{3/2}}\widehat{\Phi}_c(\pi) \right) \left(1+\frac{\alpha_u}{k^{3/2}}\widehat{\Phi}_c(\pi) \right)^{-1}\:,
\label{eq:higgsbc}
 \end{align}
 which fixes the mass $m_f$ in terms of other parameters of the theory as
 \begin{equation}
m_f
=
-\frac{v_hY}{k}\frac{\displaystyle \left[1+(\alpha_y/k^{3/2})\widehat{\Phi}_c(\pi) \right] 
N_{Q_L}N_{U_R}}
{\displaystyle 
\sqrt{1+(\alpha_u/k^{3/2})\widehat{\Phi}_c(\pi)} 
\sqrt{1+(\alpha_q/k^{3/2})\widehat{\Phi}_c(\pi)}}
\frac{G(-c_u,\alpha_u,\beta_u,\pi)}{G(-c_q,\alpha_q,\beta_q,\pi)}\, .
\label{eq:mf2vh}
\end{equation}

To derive the coupling of the radion to the zero modes, we restrict $\widehat{\Phi}$ to its background value in the action and expand the action to linear order in $\delta r=r(x)-r_c$. We then plug in the profiles for the zero modes and integrate over the extra dimension. As $r$ varies from its background value, the leading terms in the action can be written $S=S_c+\delta S$ with $S_c$ independent of $\delta r$ and $\delta S$ linear in $\delta r$. Before doing so we note that to linear order in the fluctuation of the radius $\delta r$, $\widehat{\Phi}$ satisfies 
\begin{equation}
\widehat{\Phi}(\theta)=\widehat{\Phi}_c(\theta)+\delta r\partial_{r}\widehat{\Phi}_c
=\widehat{\Phi}_c(\theta)+\delta r\left(k^{5/2}\alpha(\pi-\theta)e^{-4kr_c(\pi-\theta)}+\frac{\theta}{r_c} \partial_{\theta}\widehat{\Phi}_{\text{OR}}(\theta)\right) ,\label{eq:phivar_wrt_r}
\end{equation} 
where we have used \eqref{eq:phi-soln-massANDcubic}. We then find
\begin{align}
\delta S=&\!\int\! d^4x\!\int_0^{\pi}\!\! d\theta\delta r\,e^{-4kr_c\theta}\nonumber\\
&\bigg\{ ie^{kr_c\theta}\left[(1-3kr_c\theta)\left(1+\frac{\alpha_q}{k^{3/2}} \widehat{\Phi}_c \right) +\frac{r_c\alpha_q}{k^{3/2}}\partial_r\widehat{\Phi}_c\right] 
\!\!\left[\mathcal{Q}^\dag_R\sigma^{\mu}\overleftrightarrow{\partial_{\mu}} \mathcal{Q}_R+\mathcal{Q}^\dag_L \overline{\sigma}^{\mu}\overleftrightarrow{ \partial_{\mu}}\mathcal{Q}_L\right]\! \nonumber\\
 &+\!\left[\frac{\alpha_q}{r_ck^{3/2}}\partial_r \widehat{\Phi}_c -4k\theta\left( 1+\frac{\alpha_q}{k^{3/2}}\widehat{\Phi}_c\right) \right]\!\!\left[ \mathcal{Q}^\dag_L \overleftrightarrow{\partial_{\theta}}\mathcal{Q}_R- \mathcal{Q}^\dag_R\overleftrightarrow{\partial_{\theta}}\mathcal{Q}_L\right]\nonumber\\
&-2kc_q\left[(1-4kr_c\theta)\left(1+\frac{\beta_q}{k^{3/2}}\widehat{\Phi}_c \right) +\frac{r_c\beta_q}{k^{3/2}}\partial_r \widehat{\Phi}_c \right]\left[\mathcal{Q}^\dag_L\mathcal{Q}_R+\mathcal{Q}^\dag_R\mathcal{Q}_L \right]+\left(\mathcal{Q}\rightarrow \mathcal{U}\right)\nonumber\\
&+ \frac{v_hY}{k}\delta(\theta-\pi)\left[\frac{\alpha_y}{k^{3/2}}\partial_r \widehat{\Phi}_c -4k\theta\left(1+\frac{\alpha_y}{k^{3/2}}\widehat{\Phi}_c \right) \right]\! \left[\mathcal{Q}^{\dag}_R\mathcal{U}_L+ \mathcal{Q}^{\dag}_L\mathcal{U}_R+ \mathcal{U}_R^{\dag}\mathcal{Q}_L+\mathcal{U}^{\dag}_L\mathcal{Q}_R \right]\!\bigg\}  .
\label{eq:sint}
 \end{align}
Before inserting the profiles into $\delta S$ to derive the coupling of the 4D fields to the radion, we note that the following identities hold to first order in the small parameter $m_f/(ke^{-kr_c\pi})$:
 \begin{align}
i\left(\mathcal{Q}^\dag_R\sigma^{\mu}\overleftrightarrow{ \partial_{\mu}}\mathcal{Q}_R +\mathcal{Q}^\dag_L \overline{\sigma}^{\mu}\overleftrightarrow{\partial_{\mu}}\mathcal{Q}_L \right) =&m_f\left(u^{0\dag}_Rq_L^0+ q^{0\dag}_L u^0_R \right)\left(Q^0_RQ^0_R+Q^0_LQ^0_L \right),\\
\mathcal{Q}^\dag_L\mathcal{Q}_R+\mathcal{Q}^\dag_R\mathcal{Q}_L= &\left(u^{0\dag}_Rq_L^0+ q^{0\dag}_L u^0_R \right)Q^0_LQ^0_R,\\
\mathcal{Q}^\dag_L\overleftrightarrow{\partial_{\theta}}\mathcal{Q}_R- \mathcal{Q}^\dag_R\overleftrightarrow{\partial_{\theta}}\mathcal{Q}_L=& \left(u^{0\dag}_Rq_L^0+ q^{0\dag}_L u^0_R \right) \left(Q^0_L\partial_{\theta}Q^0_R- Q^0_R\partial_{\theta}Q^0_L \right).
\end{align}
Similar relations exist for the $\mathcal{U}$ field. The last of these can be further simplified by using the equation satisfied by the profiles, Eq.~\eqref{eq:Q0L-profile-eqn}. To take the boundary conditions into account, we write the boundary terms as $\delta$-functions in the equations of motion. A partial cancellation of terms results in the simplification 
 \begin{align}
&\left(Q^0_L\partial_{\theta}Q^0_R- Q^0_R\partial_{\theta}Q^0_L \right) =2Q^0_LQ^0_R\,k\,c_q\,r_c\frac{1+(\beta_q/k^{3/2})\widehat{\Phi}_c}{ 1+(\alpha_q/k^{3/2})\widehat{\Phi}_c}\nonumber\\
&-m_fr_ce^{kr_c\theta}\left(Q^0_RQ^0_R+Q^0_LQ^0_L \right) -\frac{v_hY}{2k} \delta(\theta-\pi) \frac{1+(\alpha_y/k^{3/2})\widehat{\Phi}_c}{ 1+(\alpha_q/k^{3/2})\widehat{\Phi}_c}
\left(Q^0_RU^0_L+Q^0_LU^0_R \right).
 \end{align}
Similarly, the term in Eq.~\eqref{eq:sint} for $\delta S$ localized 
at $\theta=\pi$ can be expressed as
 \begin{equation}
\mathcal{Q}^\dag_R\mathcal{U}_L+ \mathcal{Q}^\dag_L\mathcal{U}_R +\mathcal{U}^\dag_R\mathcal{Q}_L+ \mathcal{U}^\dag_L\mathcal{Q}_R =\left( u^{0\dag}_Rq_L^0+ q^{0\dag}_L u^0_R \right) \left(Q^0_RU^0_L+U^0_RQ^0_L \right).
\end{equation}
Inserting these relations into Eq.~\eqref{eq:sint} we find
\begin{align}
\delta S=&\int d^4x\left( u^{0\dag}_Rq_L^0+ q^{0\dag}_L u^0_R \right) 
\delta r\int_0^{\pi}d\theta e^{-4kr_c\theta} \nonumber\\
&\Bigg[
m_fe^{kr_c\theta}(1+kr_c\theta)
\left(1+\frac{\alpha_q}{k^{3/2}}\widehat{\Phi}_c \right)
\left(Q^0_RQ^0_R+Q^0_LQ^0_L\right)
\nonumber\\
&\qquad\qquad
-2kc_q
\left(1+\frac{\beta_q}{k^{3/2}}\widehat{\Phi}_c+ \frac{\beta_q-\alpha_q}{k^{3/2}+\alpha_q\widehat{\Phi}_c}r_c\partial_r \widehat{\Phi}_c \right) 
Q^0_LQ^0_R
\;\;+\;\;
\left(Q^0\rightarrow U^0\right)
\nonumber\\
&
+\frac{v_hY}{k}\delta(\theta-\pi)
\frac{\partial_r \widehat{\Phi}}{k^{3/2}} \times \nonumber \\ 
&\left\{\!
\alpha_y\!
-\!\frac{1}{2}\!
\left(\frac{\alpha_q(1+(\alpha_y/k^{3/2})\widehat{\Phi}_c)} {1+(\alpha_q/k^{3/2})\widehat{\Phi}_c} + \frac{\alpha_u(1+(\alpha_y/k^{3/2})\widehat{\Phi}_c)} {1+(\alpha_u/k^{3/2})\widehat{\Phi}_c}\right) 
\right\} 
\!\left(\!Q^0_RU^0_L+U^0_RQ^0_L \!\right)\!\Bigg].
\end{align}

To proceed further, we insert the functional form of the profiles, use the relationship between $m_f$ and $v_h$, and work to linear order in $m_f/(ke^{-kr_c\pi})$. To this order, we can drop terms like $Q^0_RQ^0_R$ and $U^0_LU^0_L$ in the above. This results in
 \begin{align}
\delta S\!=\!&\int d^4x\left( u^{0\dag}_Rq_L^0+ q^{0\dag}_L u^0_R \right) 
\delta r\,
m_f\int_0^{\pi}d\theta\Bigg\{\Bigg[
N_{Q_L}^2(1+kr_c\theta)G^2(c_q,\alpha_q,\beta_q,\theta)
\nonumber\\
&+
\frac{2c_qkr_cN_{Q_L}^2}{1+(\alpha_q/k^{3/2})\widehat{\Phi}_c} 
\left(1+\frac{\beta_q}{k^{3/2}}\widehat{\Phi}_c +\frac{\beta_q-\alpha_q }{k^{3/2}+\alpha_q\widehat{\Phi}_c}r_c\partial_r \widehat{\Phi}_c \right)
\int_0^{\theta}d\theta ' G^2(c_q,\alpha_q,\beta_q,\theta ')\Bigg]
\nonumber\\
&+\Bigg[
N_{U_R}^2(1+kr_c\theta)G^2(-c_u,\alpha_u,\beta_u,\theta)
\nonumber\\
&-
\frac{2c_ukr_cN_{U_R}^2}{1+(\alpha_u/k^{3/2})\widehat{\Phi}_c} 
\left(1+\frac{\beta_u}{k^{3/2}}\widehat{\Phi}_c +\frac{\beta_u-\alpha_u }{k^{3/2}+\alpha_u\widehat{\Phi}_c}r_c\partial_r \widehat{\Phi}_c \right)
\int_0^{\theta}d\theta ' G^2(-c_u,\alpha_u,\beta_u,\theta ')\Bigg]
\nonumber\\
&
-\delta(\theta-\pi)\frac{\partial_r \widehat{\Phi}_c}{2k^{3/2}}\!
\left( 
\frac{2\alpha_y}{1+(\alpha_y/k^{3/2})\widehat{\Phi}_c} 
-\frac{\alpha_q}{1+(\alpha_q/k^{3/2})\widehat{\Phi}_c}
-\frac{\alpha_u}{1+(\alpha_u/k^{3/2})\widehat{\Phi}_c}
\right)\!\!\!
\Bigg\}.
 \end{align}
For compactness, we write $\delta S$ as
 \begin{equation}
\delta S=-\int d^4x\left( u^{0\dag}_Rq_L^0+ q^{0\dag}_L u^0_R \right) \frac{\widetilde{\varphi}}{f}m_f\left(I_q+I_u+I_h \right)\:,
 \end{equation}
where $I_q$ ($I_u$) is associated with the term in the first (second) set of 
square brackets and $I_h$ arises from the final boundary term,
 \begin{equation}
I_h=-\frac{\widehat{\Phi}_{\text{OR}}'(\pi)}{2} \left[\frac{2\alpha_y}{k^{3/2}+\alpha_y\widehat{\Phi}_c(\pi)}- \frac{\alpha_q}{k^{3/2}+\alpha_q\widehat{\Phi}_c(\pi)}- \frac{\alpha_u}{k^{3/2}+\alpha_u\widehat{\Phi}_c(\pi)}\right],
 \end{equation}
where we have used Eq.\eqref{eq:phivar_wrt_r} and $\frac{d\phantom{kr_c\theta}}{d(kr_c\theta)}\widehat{\Phi}\equiv\widehat{\Phi} '$.

The quantity $I_q$ is related to $I_u$ by taking $c_q\rightarrow -c_u$ 
and replacing the $q$ labels with $u$ labels on $\alpha_q$ and 
$\beta_q$. This holds quite generally for the rest of this appendix, and 
so we limit our attention to $I_q$. Using the definition of $N_{Q_L}$ 
from Eq.~\eqref{eq:Qnorm} and integrating by parts, the expression for 
$I_q$ simplifies to
\begin{align}
I_q=&\frac{1}{2kr_c\pi}+ c_q+\frac{c_q}{\pi k^{3/2}}(\beta_q-\alpha_q)\int^{\pi}_0 \frac{d\theta}{1+\displaystyle\frac{\alpha_q}{k^{3/2}} \widehat{\Phi}_c}\left(\widehat{\Phi}_c +\frac{r_c\partial_r \widehat{\Phi}_c} {1+\displaystyle\frac{\alpha_q}{k^{3/2}} \widehat{\Phi}_c} \right)\nonumber\\
&+N^2_{Q_L}\frac{r_c}{\pi}\int_0^{\pi}d\theta e^{kr_c\theta(1-2c_q)} e^{-2c_qkr_c(\beta_q-\alpha_q)T(\alpha_q,\theta)}\nonumber\\
&\times \left[\theta(1-2c_q)- \frac{2c_q}{k^{3/2}}(\beta_q-\alpha_q)\int_0^{\theta}\frac{d\theta '}{1+\displaystyle\frac{\alpha_q}{k^{3/2}} \widehat{\Phi}_c}\left(\widehat{\Phi}_c +\frac{r_c\partial_r \widehat{\Phi}_c} {1+\displaystyle\frac{\alpha_q}{k^{3/2}} \widehat{\Phi}_c} \right)\right] \, ,
\label{eq:first-Iq}
\end{align}
 where we have used the fact that $\widehat{\Phi}(\theta)$ is even about 
$\theta=0$. From Eqs. \eqref{eq:phi-soln-massANDcubic} and 
\eqref{eq:phivar_wrt_r}, this expression depends on both 
$\widehat{\Phi}_{\text{OR}}$ and on $\alpha\exp[-4kr_c(\pi-\theta)]$. The effects of this second term, however, are small and can be neglected. This is 
because this term is only significant in a small region close to the IR 
brane, and so the region of integration where it has support is 
parametrically small. Therefore its contribution is suppressed by the size 
of this region, $\mathcal{O}(1/kr_c)$. Therefore, in the rest of this 
section, we drop all $\alpha\exp[-4kr_c(\pi-\theta)]$ terms and replace 
$\widehat{\Phi}_c$ with $\widehat{\Phi}_{\text{OR}}$.

Integrating Eq.~\eqref{eq:first-Iq} by parts we then find
\begin{align}
I_q=&c_q+c_q\frac{(\beta_q-\alpha_q) \widehat{\Phi}_{\text{OR}}(\pi)}{k^{3/2}+\alpha_q \widehat{\Phi}_{\text{OR}}(\pi)}\nonumber\\
&+\frac{1}{2kr_c}\left[\int_0^{\pi}d\theta e^{kr_c(\theta-\pi)(1-2c_q)} \exp\left(2c_q kr_c(\beta_q-\alpha_q)\int_{\theta}^{\pi}d\theta '\frac{\widehat{\Phi}_{\text{OR}}(\theta ')} {k^{3/2}+\alpha_q\widehat{\Phi}_{\text{OR}}(\theta ')} \right)\right]^{-1}\nonumber\\
=&c_q+c_q\frac{(\beta_q-\alpha_q)\widehat{\Phi}_{\text{OR}}(\pi)}{k^{3/2}+\alpha_q \widehat{\Phi}_{\text{OR}}(\pi)} +\frac{G^2(c_q,\alpha_q,\beta_q,\pi)}{2kr_c\displaystyle \int_0^{\pi}d\theta \,G^2(c_q,\alpha_q,\beta_q,\theta)} 
\label{eq:Iq_exact}
\end{align}
where we have used the notation of Eq. \eqref{eq:T}. This expression 
can be further simplified in the cases of phenomenological interest. 
Recall that these $G$ functions are tied to the fermion profiles in the 
$\theta$ dimension. In the unstabilized case, the fermion profiles are 
peaked towards one brane and exponentially small near the other. If 
$\widehat{\Phi} / k^{3/2}$ is large or rapidly varying, then the fermion profiles could in principle have much more complicated behavior, such as local 
extrema in the bulk. In the rest of the analysis, we will focus on the 
phenomenologically interesting case when the profiles are peaked towards 
either $\theta=0$ or $\theta=\pi$. Because the $G$ function is an 
exponential, in general when it is peaked near the one brane, it is 
exponentially small near the other. We can use this fact to simplify the 
integral in Eq.~\eqref{eq:Iq_exact}. When the fermion profile 
is peaked near $\theta=0$, we can immediately see that 
$G(c_q,\alpha_q,\beta_q,\pi)$ is exponentially suppressed, making the 
third term in Eq. \eqref{eq:Iq_exact} negligible.

When the fermion profile is peaked near $\theta=\pi$ we make the change 
of variables $\theta\to\pi-\vartheta$ in the integral in the denominator 
of the third term to obtain
 \begin{equation}
I_q=c_q+c_q\frac{(\beta_q-\alpha_q)\widehat{\Phi}_{\text{OR}}(\pi)}{k^{3/2}+\alpha_q \widehat{\Phi}_{\text{OR}}(\pi)} +\frac{1}{2kr_c}\left[\int_0^{\pi}d\vartheta e^{f(\vartheta)} \right]^{-1}
\end{equation}
where
\begin{equation}
f(\vartheta)=-kr_c\vartheta (1-2c_q)+2kr_cc_q(\beta_q-\alpha_q) \int_{\pi-\vartheta}^{\pi}d\theta\frac{\widehat{\Phi}_{\text{OR}}(\theta)} {k^{3/2}+\alpha_q\widehat{\Phi}_{\text{OR}}(\theta)}.
\end{equation}
The integral is now dominated by values of the integrand close to 
$\vartheta=0$. Now, notice that the leading terms in the Taylor series 
expansion of $f(\vartheta)$ about $\vartheta=0$ are
\begin{equation}
\vartheta\left[-kr_c(1-2c_q)+2kr_cc_q(\beta_q-\alpha_q) \frac{\widehat{\Phi}_{\text{OR}}(\pi)} {k^{3/2}+\alpha_q\widehat{\Phi}_{\text{OR}}(\pi)}\right]-\vartheta^2 \frac{k^2r_c^2c_q(\beta_q-\alpha_q)k^{3/2}\widehat{\Phi}_{\text{OR}} '(\pi)}{(k^{3/2}+\alpha_q\widehat{\Phi}_{\text{OR}}(\pi))^2}\,.
\end{equation}
Close to $\vartheta=0$ the linear term dominates and we treat the quadratic term as a small correction. We therefore write the integral as
\begin{align}
\int_0^{\pi}d\vartheta&\exp\left[kr_c\vartheta\left(-1+2c_q+2c_q(\beta_q-\alpha_q) \frac{\widehat{\Phi}_{\text{OR}}(\pi)}{k^{3/2}+\alpha_q \widehat{\Phi}_{\text{OR}}(\pi)} \right) \right]\nonumber\\
&\times\left[1- \vartheta^2 \frac{k^2r_c^2c_q(\beta_q-\alpha_q)k^{3/2}\widehat{\Phi}_{\text{OR}} '(\pi)}{(k^{3/2}+\alpha_q\widehat{\Phi}_{\text{OR}}(\pi))^2}\right]
\end{align} 
which is evaluated exactly as
\begin{equation}
\frac12(1-2 \widetilde{c}_q) \left[1-\frac{2\widetilde{c}_q(\beta_q-\alpha_q) \widehat{\Phi}_{\text{OR}}'(\pi)k^{3/2}} {(1-2\widetilde{c}_q)^2(k^{3/2}+\alpha_q\widehat{\Phi}_{\text{OR}}(\pi)) (k^{3/2}+\beta_q\widehat{\Phi}_{\text{OR}}(\pi))} \right]^{-1}
\end{equation}
were we have defined 
\begin{equation}
\widetilde{c}_q\equiv c_q+c_q\frac{(\beta_q-\alpha_q)\widehat{\Phi}_{\text{OR}}(\pi)}{k^{3/2}+\alpha_q \widehat{\Phi}_{\text{OR}}(\pi)}.\label{eq:newcq}
\end{equation}
In the limit $k^{-3/2}\widehat{\Phi} '(\pi)\ll 1$, corresponding to a light 
radion, we therefore find
\begin{equation}
I_q=\frac12+\frac{\widetilde{c}_q(\beta_q-\alpha_q) \widehat{\Phi}_{\text{OR}}'(\pi)k^{3/2}} {(1-2\widetilde{c}_q)(k^{3/2}+\alpha_q\widehat{\Phi}_{\text{OR}}(\pi)) (k^{3/2}+\beta_q\widehat{\Phi}_{\text{OR}}(\pi))}\,.
\end{equation}

Combining the two cases, we obtain
\begin{equation}
I_q=\left\{\begin{array}{lc}
\displaystyle \widetilde{c}_q & \, , \; \displaystyle G\text{ peaked at }\theta=0\\
&\\
\displaystyle \frac12+\frac{\widetilde{c}_q(\beta_q-\alpha_q) \widehat{\Phi}_{\text{OR}}'(\pi)k^{3/2}} {(1-2\widetilde{c}_q)(k^{3/2}+\alpha_q\widehat{\Phi}_{\text{OR}}(\pi)) (k^{3/2}+\beta_q\widehat{\Phi}_{\text{OR}}(\pi))} & \, , \; \displaystyle G\text{ peaked at }\theta=\pi
\end{array}\right. .
 \end{equation}

\section{Effects of Radion Stabilization on Operator Scaling Dimensions 
\label{app:b}}

In this appendix we determine how the scaling dimension 
$\Delta_{\mathcal{Q}}$ of the dual CFT operator associated with the 
fermion field $\mathcal{Q}$ is affected by the dynamics that stabilizes 
the radion. We follow closely the approach of~\cite{Contino:2004vy}. The 
central idea is to relate the bulk physics to that of a CFT by treating 
a bulk field and its boundary value as separate fields, and then 
integrating out the bulk physics.

Using Eq.~\eqref{eq:veilbein} and Eq.~\eqref{eq:5dgamma} we begin with the 
fermion action \begin{align}
 S&=\int d^4x\int_0^{\pi}d\theta \Bigg\{ \nonumber\\
&\frac{1}{2} \bigg[ ir_ce^{-3kr_c\theta}\!\left(\! \mathcal{Q}^{\dag}_R\sigma^{\mu}\overleftrightarrow{\partial_{\mu}}\mathcal{Q}_R +\mathcal{Q}^{\dag}_L\sigma^{\mu}\overleftrightarrow{\partial_{\mu}}\mathcal{Q}_L\!\right) +e^{-4kr_c\theta}\!\left(\!\mathcal{Q}^{\dag}_L\overleftrightarrow{\partial_{\theta}}\mathcal{Q}_R- \mathcal{Q}_R^{\dag}\overleftrightarrow{\partial_{\theta}}\mathcal{Q}_L \!\right)\!\bigg]\!\! \left(1+\frac{\alpha_q}{k^{3/2}}\widehat{\Phi}_c\right) \nonumber\\
&\;\;\;\;\;\;\;\;  -kc_qr_ce^{-4kr_c\theta}\left[\mathcal{Q}_L^{\dag}\mathcal{Q}_R +\mathcal{Q}^{\dag}_R\mathcal{Q}_L \right] \left( 1+\frac{\beta_q}{k^{3/2}}\widehat{\Phi}_c\right)\Bigg\}
\end{align}
where 
$\overleftrightarrow{\partial_{\mu}}\equiv\overrightarrow{\partial_\mu}-\overleftarrow{\partial_\mu}$ 
and $\widehat{\Phi}_c(\theta)$ is the VEV of the GW scalar. In this appendix we focus on the $\mathcal{Q}$ field. The end result can be mapped to the $\mathcal{U}$ by simply taking $c_q\rightarrow -c_u$ while changing all other $q$ labels to $u$ labels. The scaling dimension is associated with physics above the conformal symmetry breaking scale, and so in this appendix we can safely 
ignore details of the IR brane dynamics such as couplings to the Higgs.

In Appendix~\ref{app:a}, we took the $\mathcal{Q}_L$ field and $\widehat{\Phi}_c$ to be even about $\theta=0$ and $\theta = \pi$, and $\mathcal{Q}_R$ to be odd. In this appendix, solely for the purpose of determining the scaling dimensions of bulk fields, we relax those restrictions. We now associate the value of $\mathcal{Q}_L$ on the UV ($\theta$=0) brane with 
the source $q_s$ for some fermionic operator $\mathcal{O}_{\mathcal{Q}}$ 
in the CFT on the boundary with scaling dimension $\Delta_{{\mathcal{Q}}}$. Specifically,
 \begin{equation}
\mathcal{Q}_L(x,\theta)\Big\vert_{\theta =0}=q_s(x),\;\Rightarrow\; \mathcal{L}_{\text{CFT}}\supset q_s\mathcal{O}_{\mathcal{Q}}.
\label{eq:correspond}
 \end{equation}
 This function is fixed, or $\delta\mathcal{Q}_L=0$, on the UV boundary. 
Because the equation of motion for fermions is first order, we cannot 
fix the boundary conditions for both chiralities of $\mathcal{Q}$ so we 
leave $\mathcal{Q}_R$ free to vary on the boundary.

When we take the variation of the action,\footnote{We treat $\widehat{\Phi}_c$ as a background field, so $\delta\widehat{\Phi}_c=0$.} we generate the equations of motion such as Eq.~\eqref{eq:QLeq} and a boundary term such as led to 
Eq.~\eqref{eq:QLbc}. The total boundary term is
 \begin{equation} 
\delta S\supset \frac{1}{2}\int d^4xe^{-4kr_c\theta}\left( 
1+\frac{\alpha_q}{k^{3/2}}\widehat{\Phi}_c\right) \left[\mathcal{Q}_L^{\dag}\delta\mathcal{Q}_R 
-\delta\mathcal{Q}_L^{\dag}\mathcal{Q}_R- \mathcal{Q}_R^{\dag}\delta\mathcal{Q}_L+ 
\delta\mathcal{Q}_R^{\dag}\mathcal{Q}_L \right]\Big\vert_0^{\pi}. 
 \end{equation} 
Now we choose $\mathcal{Q}_L|_{\pi}=0$ to eliminates the boundary term at $\theta=\pi$.\footnote{It would also be consistent to choose $\mathcal{Q}_R|_{\pi}=0$. This has no effect on the final result.} 
The UV boundary, where $\delta\mathcal{Q}_L=0$, remains because 
$\mathcal{Q}_L\neq 0$ and $\delta\mathcal{Q}_R\neq 0$. Thus, in order 
for $\delta S=0$ to hold we must add a term on the UV boundary to cancel 
this remainder. This term is
 \begin{equation} 
S_4=\frac{1}{2}\int_{\text{UV}} 
d^4x \left(1+\frac{\alpha_q}{k^{3/2}}\widehat{\Phi}_c \right)\left(\mathcal{Q}^{\dag}_L\mathcal{Q}_R 
+\mathcal{Q}_R^{\dag}\mathcal{Q}_L \right) 
\end{equation} 
where all the fields are evaluated at the UV brane.

Because $\delta\mathcal{Q}_L=0$ on the UV brane, we can also add to the 
boundary Lagrangian any term $\mathcal{L}_{\text{UV}}$ which is only a 
function of $\mathcal{Q}_L$ without changing the equations of motion. 
For instance
 \begin{equation}
S_{\text{UV}}=\int_{\text{UV}} d^4x\mathcal{L}_{\text{UV}}= \int_{\text{UV}} d^4x\left[\left(\widehat{\beta}_q+\frac{\widehat{\alpha}_{q}} {k^{3/2}}\widehat{\Phi}_c\right)i\mathcal{Q}^{\dag}_L\slashed{\partial}\mathcal{Q}_L+\cdots \right]
\label{eq:brane-lag}
\end{equation}
where $\widehat{\beta}_q$ and $\widehat{\alpha}_{q}$ are arbitrary coefficients. We are now ready to integrate out the bulk by substituting the solutions to the 5D equations of motion back into the action. By design, the bulk action vanishes when the variation vanishes, so we are left with only the UV boundary terms.

It is useful to Fourier transform the 4D coordinates of the 5D fields and 
parametrize their $\theta$ dependence by
 \begin{equation}
\begin{array}{cc}
\displaystyle \mathcal{Q}_L(p,\theta)=\frac{f_L(p,\theta)}{f_L(p,0)}q_s(p), \quad &\displaystyle \mathcal{Q}_R(p,\theta)=\frac{f_R(p,\theta)}{f_R(p,0)}q_R(p)
\end{array}
 \end{equation} 
 where we have made the definition $\mathcal{Q}_R(p,0)=q_R(p)$. The 4D 
fermions $q_R$ and $q_s$ are related by the Dirac equation, and we can 
fix the relative normalization by taking
\begin{equation}
\slashed{p}q_s=p\frac{f_L(p,0)}{f_R(p,0)}q_R.
\end{equation} 
The bulk equations of motion for $\mathcal{Q}_L$ and $\mathcal{Q}_R$ 
then imply
\begin{equation}
\partial_{\theta}f_{L,R}\mp kr_c\left( \frac{p}{k}e^{kr_c\theta}f_{R,L}-2 f_{L,R} \pm c_q\frac{k^{3/2}+\beta_q\widehat{\Phi}_c}{k^{3/2}+\alpha_q\widehat{\Phi}_c}f_{L,R} +\frac{\alpha_q}{2} \frac{\displaystyle \widehat{\Phi}_c '}{k^{3/2}+ \alpha_q\widehat{\Phi}_c}f_{L,R}\right) =0\label{eq:frleqs}
\end{equation}
where the first and second labels on $f_{i,j}$ correspond to the upper 
and lower signs respectively and $\widehat{\Phi}_c '\equiv \frac{d\phantom{kr_c\theta}}{d(kr_c\theta)} \widehat{\Phi}_c$.

We now turn to the boundary action for $q_s$. After rescaling so that 
the kinetic term in $S_{UV}$ is canonically normalized, the action is 
\begin{equation} 
S= \int \frac{d^4p}{(2\pi)^4}\left[q_s^{\dag}\slashed{p}q_s+\cdots \right] 
+\hat{\zeta}_q\, q_s^{\dag}\Sigma(p)q_s 
\label{eq:sholog} 
 \end{equation} 
where $\hat{\zeta}_q$ is a a normalization that is independent of $p$ 
and we have defined 
 \begin{equation} 
\Sigma(p)=\frac{\slashed{p}}{p}\frac{f_R(p,0)}{f_L(p,0)}. 
\label{eq:sigma-def} 
 \end{equation} 
 This quantity $\Sigma(p)$ determines the scaling dimension of the 
operator $\mathcal{O}_{\mathcal{Q}}$ defined in Eq.~\eqref{eq:correspond}. 
This follows from the holographic principle which associates the 
generating function of a CFT
 \begin{equation} 
\mathcal{Z}[q_s]=\int\mathcal{D}\phi_{\text{CFT}}e^{iS_{\text{CFT}}
+\int q_s\mathcal{O}_{\mathcal{Q}}+\text{h.c.}} 
 \end{equation} 
 with the AdS partition function
 \begin{equation} 
\mathcal{Z}[q_s]=\int_{q_s}\mathcal{D}\mathcal{Q} e^{i S[\mathcal{Q}]}= 
e^{i S_{\text{eff}}[q_s]} 
 \end{equation} 
 where the subscript on the integral means that the integration is to be 
performed subject to the condition that $\mathcal{Q}$ takes on the value 
$q_s$ on the UV boundary. In general $S_{\text{eff}}[q_s]$ represents a 
nonlocal action in 4D for the source field $q_s$. In its original 
incarnation, the AdS/CFT correspondence applied to the scenario when the 
UV boundary, where the source field resides, corresponded to the boundary 
of AdS space. However, since different positions of the UV brane 
correspond simply to different choices of the cutoff scale, and are 
therefore related by RG transformations, the holographic principle can 
also be applied to the situation when the UV brane sits at an arbitrary 
location in the bulk. In general, it is also possible to promote 
$q_s(x)$ to a dynamical field. The $\mathcal{L}_{\text{UV}}$ term we can 
add to the UV brane in Eq.~\eqref{eq:brane-lag} exactly captures this 
freedom.

Since the scaling dimension of an operator is associated with physics 
above the conformal symmetry breaking scale, we work in the limit of the 
IR brane being far away. This is done by taking the limit $p \gg k \exp(-k \pi r_c)$, where $p$ represents the momentum scales being 
probed. It is likewise convenient to work in the limit that the UV brane 
is also far away, so that $p \ll k$. The reason is that the hard 
momentum cutoff associated with the presence of the UV brane constitutes 
an explicit violation of conformal symmetry by the regulator. When 
working at momenta well below the cutoff scale, spurious effects 
associated with the regulator are suppressed. In this limit, for 
instance, the correlator
 \begin{equation}
\langle \overline{\mathcal{O}}_{\mathcal{Q}}\mathcal{O}_{\mathcal{Q}}\rangle \simeq \int \frac{d^4p}{(2\pi)^4}e^{-ix\cdot p}\frac{\delta^2S_{\text{eff}}}{\delta q_s^{\dag} \delta q_s}
 \end{equation}
 has dimension $2\Delta_{{\mathcal{Q}}}$. This allows us to relate
 \begin{equation}
\lim_{\substack{kr_c\to \infty\\ p/k\to 0}}\left(\Sigma(p)+\text{counterterms} \right)
 \end{equation}
 to $\Delta_{{\mathcal{Q}}}$. The counterterms are included because 
divergent terms in $\Sigma(p)$ which are local, and hence analytic, are 
renormalized by local counterterms. This implies that the leading 
nonanlaytic term in $\Sigma(p)$ gives the dimension of 
$\mathcal{O}_{\mathcal{Q}}$; specifically the leading nonanalytic term 
goes like~\cite{Contino:2004vy}
 \begin{equation}
\lim_{\substack{kr_c\to \infty\\ p/k\to 0}}\left( \Sigma(p)+\text{counterterms} \right)\propto\slashed{p}p^{2\Delta_{{\mathcal{Q}}}-5}. \label{eq:nonanalyticdim}
 \end{equation}
 Therefore once we compute $\Sigma(p)$, it will give us the scaling 
dimension of the dual operator.

The bulk RS metric possesses an isometry under shifts in the extra 
dimensional coordinate $\theta$, when combined with a rescaling of the 
4D coordinates $x^{\mu}$. This isometry corresponds to the symmetry 
under scale invariance of the dual 4D theory. After the introduction of 
the stabilization mechanism, the isometry of the bulk 5D metric is no 
longer exact. In the dual description, the scale invariance of the 4D 
theory is now explicitly broken, and the scaling dimensions of operators 
are no longer strictly defined. However, the scenario we are interested 
in is one where the dilaton is light as a consequence of the fact that 
the operator that breaks the symmetry is close to marginal, and so the 
theory is approximately conformally invariant at all scales. In this 
limit, the scaling behavior of operators only changes very slowly as a 
function of the renormalization scale. Therefore we can continue to 
associate each operator with an approximate scaling dimension that 
changes very slowly with the renormalization scale. From the holographic 
perspective, the scaling dimension is dual to a function of the 
parameters of the 5D theory that changes with the extra dimensional 
coordinate $\theta$, but only very slowly.

Ultimately, we are interested in comparing with the dual 
picture~\cite{Chacko:2012sy}, where the dilaton couplings are related to 
the scaling dimensions of operators evaluated near the scale where 
conformal symmetry is spontaneously broken. Therefore, we need to relate 
the scaling dimensions to 5D parameters evaluated near the IR brane. 
This is challenging because at $\theta=\pi$ there is a phase transition 
where a boundary layer forms. Therefore we need to be careful in 
taking the limit approaching the IR brane, and we use the specific 
procedure described below.

We first modify the analysis above to compute the scaling dimension, 
$\Delta_\mathcal{Q}(\theta_0)$, in the neighborhood of an arbitrary point 
$\theta = \theta_0$ in the bulk. We imagine that there is a UV brane at 
$\theta=\theta_0$ such that
 \begin{equation}
\mathcal{Q}_L(x,\theta)\Big\vert_{\theta =\theta_0}=q_s(x|\theta_0),\;\Rightarrow\; \mathcal{L}_{\text{CFT}}\supset q_s(x| \theta_0)\mathcal{O}_{\mathcal{Q}}.
 \end{equation}
 We can now follow the analysis above to get the equations analogous to 
Eqs.~\eqref{eq:sholog} and~\eqref{eq:sigma-def}, namely
 \begin{equation}
S= \int \frac{d^4p}{(2\pi)^4}\left[q_s^{\dag}(\theta_0)\slashed{p}q_s(\theta_0)+\cdots \right] +\hat{\zeta}_q\, q_s^{\dag}(\theta_0) \Sigma_{\theta_0}(p)q_s(\theta_0)
 \end{equation}  
with the definition
 \begin{equation}
\Sigma_{\theta_0}(p)=\frac{\slashed{p}}{p}\frac{f_R(p,\theta_0)}{f_L(p,\theta_0)}. \label{eq:2pttheta0}
 \end{equation}
 which is then related to the scaling dimension 
$\Delta_\mathcal{Q}(\theta_0)$ in the same way as before.

We now turn to calculating $\Sigma(p)$ and its generalization $\Sigma_{\theta_0}(p)$. From Eq.~\eqref{eq:frleqs} we find
\begin{equation}
f_R=\frac{e^{-kr_c\theta}}{pr_c}\left[ \partial_{\theta}f_L-2kr_cf_L+ \frac{kr_c}{2}\frac{\alpha_q\widehat{\Phi}_c '}{k^{3/2}+\alpha_q\widehat{\Phi}_c}f_L +kr_cc_q\frac{ k^{3/2}+\beta_q\widehat{\Phi}_c}{ k^{3/2}+\alpha_q\widehat{\Phi}_c}f_L \label{eq:frfull}\right].
\end{equation}
Substituting this back into the companion relation for $f_L$ we find
\begin{align}
0=&\partial_{\theta}^2f_L-kr_c\left( 5-C_1 \right)\partial_{\theta}f_L \nonumber\\
&+f_L(kr_c)^2\left\{ \frac{p^2}{k^2}e^{2kr_c\theta}+6-c_q(1+c_q)- \frac{5}{2}C_1-\frac14 C_1^2+C_2-c_qC_3 \right\}
\label{eq:fLfullode}
\end{align}
where
\begin{align}
C_1&=\frac{\alpha_q \widehat{\Phi}_c'}{k^{3/2}+\alpha_q\widehat{\Phi}_c},\\
C_2&=\frac12\frac{\alpha_q\widehat{\Phi}_c''}{k^{3/2}+\alpha_q\widehat{\Phi}_c}+ \frac{c_q (\beta_q-\alpha_q)\widehat{\Phi}_c'}{\displaystyle k^{3/2}\left(1+\frac{\alpha_q}{k^{3/2}}\widehat{\Phi}\right)^2},\\
C_3&=\frac{(\beta_q-\alpha_q)\widehat{\Phi}_c}{k^{3/2}+\alpha_q\widehat{\Phi}_c}\left[ 1+2c_q +\frac{c_q(\beta_q-\alpha_q)\widehat{\Phi}_c}{k^{3/2}+\alpha_q\widehat{\Phi}_c}\right].
\end{align}

The $C_i$ above conveniently encapsulate the $\widehat{\Phi}_c$ dependence in the differential equation for $f_L$. We are interested in theories where the 
dilaton is light, which correspond to scenarios where $\widehat{\Phi}_c$ is a 
slowly varying function of $\theta$. In this limit we can solve the 
differential equation by making a WKB approximation, treating the $C_i$ 
as constants independent of $\theta$. With this assumption the solution 
to Eq.~\eqref{eq:fLfullode} takes the form
 \begin{equation}
f_L(\theta)=e^{kr_c(5-C_1)\theta/2}\left[A_1J_{n(\theta)}\left(\frac{p}{k}e^{kr_c\theta} \right)+A_2J_{-n(\theta)}\left(\frac{p}{k}e^{kr_c\theta} \right) \right]
 \end{equation}
 where
 \begin{equation}
n(\theta)=\left|\left(c_q+\frac12\right)\sqrt{1+\frac{2C_1^2-4C_2+4c_qC_3} {(1+2c_q)^2}}\right|.
 \end{equation}

We can check if this function solves the ODE Eq.~\eqref{eq:fLfullode}. In so doing we employ an identity, which can be most easily seen from the infinite series definition of the Bessel function:
\begin{equation}
\frac{d}{dx}J_{n(x)}(x)=J'_{n(x)}(x)+\frac{dn(x)}{dx}\frac{d}{dn}J_{n}(x)
\end{equation} 
where $J'_n$ is the usual derivative of the Bessel function with 
respect to its argument. The first term of the above is like those 
terms that appear when $n$ has no dependence on $x$ and will satisfy the 
ODE. The second term leads to terms that will not satisfy the ODE. These 
terms, however, are proportional to $\frac{dn}{d\theta}$ which is in 
turn proportional to $\widehat{\Phi}_c'$ or higher order derivatives of $\widehat{\Phi}_c$ with no compensating large factors. Therefore, our solution for $f_L$ is valid up to small corrections proportional to derivatives of the slowly varying function $\widehat{\Phi}_c$.

Then, to be consistent, we must drop all $\widehat{\Phi}_c'$ terms in the $C_i$. In this limit $C_1$ and $C_2$ vanish, leaving us with the solution
 \begin{equation}
f_L(\theta)=e^{\tfrac52 kr_c\theta}\left[A_1J_{n(\theta)}\left(\frac{p}{k}e^{kr_c\theta} \right)+A_2J_{-n(\theta)}\left(\frac{p}{k}e^{kr_c\theta} \right) \right]\label{eq:flgensol}
 \end{equation}
 where
 \begin{align}
n(\theta)=&\left|\left(c_q+\frac12\right)\sqrt{1+\frac{4c_qC_3(\theta)} {(1+2c_q)^2}}\right|=\left| \frac12+c_q+ \frac{c_q(\beta_q-\alpha_q)\widehat{\Phi}_c(\theta)} {k^{3/2}+\alpha_q\widehat{\Phi}_c(\theta)}\right|.\label{eq:fulln}
 \end{align}

Now satisfied that Eq.~\eqref{eq:flgensol} solves the ODE we enforce the IR 
brane boundary condition $f_L(\pi)=0$. This yields
 \begin{equation}
f_L(\theta)\!=\!N_Le^{\frac{5}{2}kr_c\theta}\!\left[J_{-n(\pi)}\!\left(\frac{p}{k}e^{kr_c\pi}\!\right) J_{n(\theta)}\!\left(\frac{p}{k}e^{kr_c\theta}\!\right) -J_{n(\pi)}\!\left(\frac{p}{k}e^{kr_c\pi}\!\right)J_{-n(\theta)} \!\left(\frac{p}{k}e^{kr_c\theta}\! \right)\! \right],
 \end{equation}
with $N_L$ a UV dependent normalization constant. We also find
\begin{equation}
f_R(\theta)\!=\!N_Le^{\frac52 kr_c\theta}\!\left[J_{-n(\pi)}\!\left(\frac{p}{k}e^{kr_c\pi}\!\right) \!J_{n(\theta)-1}\!\left(\frac{p}{k}e^{kr_c\theta}\!\right) +J_{n(\theta)}\!\left(\frac{p}{k}e^{kr_c\pi}\!\right)\! J_{1-n(\theta)}\!\left(\frac{p}{k}e^{kr_c\theta}\!\right)\! \right] .
\end{equation} 

With $f_L$ and $f_R$ in hand, for slowly varying $\widehat{\Phi}_c$, we evaluate 
the two point correlator on the UV brane at $\theta_0$. Using 
\eqref{eq:2pttheta0} we find
 \begin{equation}
\Sigma_{\theta_0}(p)=\frac{\slashed{p}}{p}\frac{\displaystyle J_{-n(\pi)}\!\left(\frac{p}{k}e^{kr_c\pi}\!\right) J_{n(\theta_0)-1}\!\left(\frac{p}{k}e^{kr_c\theta_0}\! \right) 
+J_{n(\pi)}\!\left(\frac{p}{k}e^{kr_c\pi}\!\right)J_{1-n(\theta_0)}\!\left(\frac{p}{k} e^{kr_c\theta_0}\!\right) }{\displaystyle J_{-n(\pi)}\!\left(\frac{p}{k}e^{kr_c\pi}\!\right) J_{n(\theta_0)}\!\left(\frac{p}{k}e^{kr_c\theta_0}\!\right)   -J_{n(\pi)}\!\left(\frac{p}{k}e^{kr_c\pi}\!\right)J_{-n(\theta_0)} \!\left(\frac{p}{k}e^{kr_c\theta_0}\! \right) }.
\label{eq:sigma}
\end{equation}

We wish to express the two point function as a power series in $p/k$ to 
determine the scaling dimension of $\mathcal{O}$. To suppress effects 
associated with spontaneous conformal symmetry breaking we work in the 
limit that the IR brane is far away by choosing $p$ such that 
$\frac{p}{k}e^{kr_c\pi}\gg 1$. In order to avoid spurious conformal 
symmetry violating effects associated with the regulator, we must also 
stay away from the UV brane by choosing $p$ such that 
$\frac{p}{k}e^{kr_c\theta_0} \ll 1$. In this limit we can employ the 
small Bessel expansion for the terms with $\theta_0$ in 
Eq.~\eqref{eq:sigma}. We also Wick rotate the momenta to tame the 
oscillations of the Bessel functions. Using the asymptotic expansions 
of the Bessel functions for both small and large argument we obtain a 
result of the form
 \begin{equation}
\lim_{kr_c\rightarrow\infty} \Sigma_{\theta_0}(p)=\frac{\slashed{p}}{k}\left[a_1(\theta_0)+a_2(\theta_0)\left(\frac{p}{k} \right)^2+\cdots+b_1(\theta_0)\left(\frac{p}{k} \right)^{|2n(\theta_0)-2|}+\cdots \right]
 \end{equation}
 where the $a_{i}$ are the coefficients of analytic terms and the $b_i$ 
those of the nonanalytic terms. The coefficients depend on the 
location of the extra dimension, $\theta_0$, but they are independent of 
$p$. Comparing the power of the $b_1$ term to Eq.~\eqref{eq:nonanalyticdim} 
we immediately find that for $n>1$ ($c_q > 1/2$ at leading order)
 \begin{equation}
\Delta_{{\mathcal{Q}}}(\theta_0)=\frac{3}{2}+n(\theta_0)
\label{eq:scaling}
\end{equation}
A similar expression can be derived for $n<1$. 

We are interested in the scaling dimensions just above the conformal 
symmetry breaking scale, which corresponds to the region just outside 
the boundary layer near the IR brane. This corresponds to 
 \begin{equation}
\theta_0=\pi-\frac{x}{kr_c}.\label{eq:pr-relation}
 \end{equation}
 where $x$ is a number of order a few. We must check that the 
approximations that led to Eq.~\eqref{eq:scaling} continue to remain 
valid this close to the IR brane. In order for the asymptotic forms of 
the Bessel functions to be applicable, $p$ must be chosen to 
simultaneously satisfy
 \begin{equation}
\begin{array}{cc}
\displaystyle \frac{p}{k}e^{kr_c\pi}\gg 1, &\displaystyle\frac{p}{k}e^{kr_c\pi}e^{-x}\ll 1.
 \end{array}
\end{equation}
 These conditions can indeed be satisfied provided $e^{-x}\ll 1$, which 
corresponds to $x$ of order a few.

We now evaluate $n(\theta_0)$ in this limit. The $\theta_0$ dependence 
of $n(\theta_0)$ comes from the outer region GW solution 
Eq.~\eqref{eq:phi-soln-massANDcubic} $\widehat{\Phi}_{\text{OR}}(\theta_0)$. In 
the limit of large $kr_c$ we find
\begin{equation}
\widehat{\Phi}_{\text{OR}}(\theta_0)=\widehat{\Phi}_{\text{OR}} (\pi)-x\widehat{\Phi}_{\text{OR}}'(\pi)+\ldots
\end{equation}
Because we have been dropping all $\widehat{\Phi}_{\text{OR}}'$ to obtain the Bessel function solution \eqref{eq:flgensol} we must also drop the second and higher terms in the expansion above. We are left with $\widehat{\Phi}_{\text{OR}}(\theta_0)=\widehat{\Phi}_{\text{OR}}(\pi)$. Therefore, 
 \begin{align}
\Delta_{{\mathcal{Q}}}\Big\vert_{\rm IR} &= \frac{3}{2}+n(\pi)=\frac32+\left|\frac12 +c_q+\frac{c_q(\beta_q-\alpha_q)\widehat{\Phi}_{\text{OR}}(\pi)} {k^{3/2}+\alpha_q\widehat{\Phi}_{\text{OR}}(\pi)} \right|\,+\mathcal{O}\left(k^{-3/2}\widehat{\Phi}_{\text{OR}} '\right) .
\end{align}

\section{Naive Dimensional Analysis Estimation of Parameters}
\label{app:c}

In this appendix we estimate the sizes of the various parameters in the 
theory, using the methods of naive dimensional analysis 
(NDA)~\cite{Manohar:1983md,Georgi:1986kr} as generalized to higher 
dimensions~\cite{Chacko:1999hg}. A more detailed explanation of some of 
these estimates may be found in Appendix~C of~\cite{Chacko:2013dra}. The 
underlying philosophy behind NDA estimates is that in a strongly coupled 
theory, the radiative corrections to any process are expected to be 
comparable at every loop order. Since holography relates the 
interactions in the bulk and on the IR brane to the dynamics of a 
strongly coupled CFT, we expect that NDA will offer a guide to the sizes 
of the parameters in these regions. The dynamics on the UV brane, on the 
other hand, is associated with the interactions of states external to 
the CFT. Therefore, we do not expect that NDA will offer a useful guide 
to the sizes of parameters on this brane.

Following~\cite{Chacko:1999hg} we can write the $D$-dimensional Lagrangian 
of a strongly coupled theory as
\begin{equation}
{\cal L}_{D}  \sim \frac{N \Lambda^{D}}{\ell_{D}} \overline{{\cal L}}(\overline{\Phi}, \partial/\Lambda)\;,
\label{eq:NDA_Lag}
\end{equation} 
Here $\overline{\Phi}$ represents the fields in the theory normalized so as to be 
dimensionless, $\Lambda$ is the cutoff of the theory, and $N$ is the number of states going around the loops. The loop factor, which comes from integrating over $D$ dimensional phase space, is given in four dimensions by $\ell_4 = 16\pi^2$, while in five dimensions it is given by $\ell_5 = 24 \pi^3$. All parameters in $\overline{\cal L}$ are dimensionless and taken to be $\mathcal{O}(1)$. Rescaling the fields so that kinetic terms are canonically normalized then gives all Lagrangian parameters in terms of the cutoff, the loop factor, and the number of states participating in the correcting loops.

We begin by analyzing the gravity Lagrangian, ${\cal L} \sim 2 M_5^3 
{\cal R}$. The above prescription allows us to relate the cutoff 
$\Lambda_{\text{IR}}$ to the 5-dimensional Planck mass $M_5$,
\begin{equation}
\Lambda_{\text{IR}} \sim \left(\frac{\ell_5}{N} \right)^{1/3} M_5 \; .
\end{equation}
 We can also estimate the size of the bulk cosmological constant 
$\Lambda_b$ that would be radiatively generated by the strong dynamics, 
\begin{equation}
\Lambda_b \sim \frac{N\Lambda_{\text{IR}}^5}{\ell_5} \sim \left(\frac{\ell_5}{N} \right)^{2/3} M_5^5 \; .
\end{equation}
Einstein's equations then allow us to estimate the natural size of the 
curvature $k$, in units of the cutoff. From Eq.~\eqref{eq:def_k} we 
obtain,
 \begin{equation}
k=\sqrt{\frac{-\Lambda_b}{24M_5^3}} \sim \frac{\Lambda_{\text{IR}}}{\sqrt{24}} \; ,
 \end{equation}
from which we can express
\begin{equation}
\left(\frac{k}{M_5}\right)^3 \sim 24^{-3/2}\frac{\ell_5}{N} \sim \frac{6}{N} \; .
\end{equation}

\subsection*{GW Scalar Potential}

Next we analyze the bulk potential for the GW field. The potential in Eq.~\eqref{eq:bulk-pot} is parametrized by $\epsilon=m^2/4k^2$ and $\xi=\eta v/8\sqrt{k}$. The bulk mass of the GW scalar is estimated to be simply $m^2 \sim \Lambda_{\text{IR}}^2$, but we need the bulk mass to be small in order for the size of the extra dimension to be stabilized at a large value~\cite{Goldberger:1999uk}. Therefore, we take the NDA estimate to be an upper bound:
\begin{equation}
\epsilon \lesssim 6\; .
\end{equation}
The bulk cubic is estimated as 
\begin{equation}
\eta \sim \sqrt{\frac{\ell_5 \Lambda_{\text{IR}}}{N}}\; .
\end{equation}
If we use NDA on the UV brane, we can estimate $v\sim 0.4$, but the dynamics of the UV brane is weakly coupled, so $v$ is expected to be smaller than its NDA value. Furthermore, we need $v$ to be small in order to have an approximately conformal dual because it corresponds to explicit breaking of the CFT. Putting it all together, we get an NDA upper bound of 
\begin{equation}
\xi\lesssim \frac{3}{\sqrt{N}}\; .
\end{equation} 
Finally, we can now estimate the size of the VEV $\widehat{\Phi}$ on the IR brane by looking at the IR brane potential parameter $\alpha$. Using the NDA prescription on the 4D brane we find that
\begin{equation}
\frac{\widehat{\Phi}(\pi)}{k^{3/2}} \simeq \frac{\alpha}{4} \sim \frac{\sqrt{N\ell_5}}{8\,\ell_4}\left( \frac{\Lambda_{\text{IR}}}{k}\right)^{5/2} \sim 1.1\sqrt{N} \, .
\end{equation}

\subsection*{Couplings of SM Fields}

In order to estimate the size of the gauge couplings in 
Eq.~\eqref{eq:gauge-kt-bulkbrane}, we work in a convention where the 
gauge field is treated on the same footing as a spacetime derivative so 
that the gauge covariant derivative is $D_\mu = \partial_\mu-iA_\mu$. 
This allows us to generalize Eq.~\eqref{eq:NDA_Lag} to ${\cal L}_{D} 
\sim \frac{N \Lambda^{D}}{\ell_{D}} \overline{{\cal 
L}}(\overline{\Phi}, \partial/\Lambda,A/\Lambda)$, where $\Phi$ represents 
the nongauge fields and $A$ the gauge fields. We can then estimate the 
size of the visible brane gauge coupling in terms of the 4D loop factor,
\begin{equation}
g_{\text{IR}} \sim \frac{4\pi}{\sqrt{N}}  \; .
\end{equation}
The bulk gauge coupling is dimensionful, and we can estimate the 
following useful combination
\begin{equation}
g_5^2 k \sim \sqrt{24}\frac{\pi^3}{N} \; .
\end{equation}
The IR coupling $g_{\text{IR}}$ and bulk gauge coupling $g_5$ are 
expected to be of order their NDA values, because they are associated 
with the strong dynamics. The UV coupling $g_{\text{UV}}$, on the other 
hand, is associated with physics external to the strong dynamics, so it 
can naturally be smaller than its NDA value.

We now estimate the couplings of the GW field to SM-like fields. We 
begin with the coupling to gauge bosons as in 
Eq.~\eqref{eq:GW-bulkGB-int} which is given schematically by
\begin{equation}
\frac{\Phi}{k^{3/2}} \left\{  \frac{\beta_{\text{UV}}}{4g_{\text{UV}}^2} \delta(\theta) + \frac{\beta}{4g_5^2} + \frac{\beta_{\text{IR}}}{4g_{\text{IR}^2}} \delta(\theta-\pi) \right\} F^2 \; .
\end{equation}
Because the gauge coupling is already scaled out of the definition of 
$\beta$, all one needs to do to estimate its size is rescale $\Phi$ so 
it is canonically normalized using the prescription of 
Eq.~\eqref{eq:NDA_Lag}. Therefore we find that
\begin{equation}
\beta \sim \sqrt{\frac{\ell_5}{N}} \left(\frac{k}{\Lambda_{\text{IR}}} \right)^{3/2} \sim \frac{2.5}{\sqrt{N}} \; .
\label{eq:beta-NDA}
\end{equation}
To estimate the IR brane coupling to bulk fields $\beta_{\text{IR}}$, we 
note that $F^2$ is normalized as a 5D operator, while in the NDA 
prescription the brane operator is multiplied by the 4D loop factor. 
Therefore, we find that
\begin{equation}
\beta_{\text{IR}} \sim \frac{\ell_5^{3/2}}{\ell_4\sqrt{N}}\left(\frac{k}{\Lambda_{\text{IR}}} \right)^{5/2} \sim \frac{2.4}{\sqrt{N}} \; ,
\end{equation}
which is numerically similar to the estimate for the bulk coupling 
$\beta$. The size of $\beta_{\text{UV}}$ is not correlated with the NDA 
estimate.

The GW scalar can also couple to the Higgs kinetic term on the IR brane 
as in Eq.~\eqref{betaWm},
\begin{equation}
\beta_W \frac{\Phi}{k^{3/2}}\:\delta(\theta-\pi)\: (\mathcal{D}^\mu H)^\dagger(\mathcal{D}_\mu H)\; .
\end{equation}
An estimate of the NDA size of $\beta_W$ yields a result similar to 
that of the gauge kinetic term on the IR brane, $\sqrt{N}\beta_W\sim 
\sqrt{N}\beta_{\text{IR}}\sim 2.4$. The coupling of the GW field to the fermions is given 
in Eq.~\eqref{eq:bulk-fermions-GW-coupling},
\begin{equation}
\:\frac{\Phi}{k^{3/2}}
\Bigg[\left(
\alpha_q\:
\frac{i}{2}e_a^M
\overline{\mathcal{Q}}\Gamma^a\overleftrightarrow{\partial_M}\mathcal{Q}
-\beta_q\:kc_q \overline{\mathcal{Q}}\mathcal{Q}
 \right)
+
\delta(\theta-\pi)\alpha_y
\left(
\frac{Y}{k}\overline{\mathcal{Q}}H\mathcal{U}+\text{h.c.}
\right)
\Bigg]\: .
\end{equation}
We have only shown $\mathcal{Q}$, but the generalization to other 
fermions is clear. We find that $\sqrt{N}\alpha_y \sim \sqrt{N}\beta_v \sim 2.4$, and 
that $\sqrt{N}\alpha_{q} \sim \sqrt{N}\beta \sim 2.5$. Here we are assuming that the coupling to the GW scalar does not break the SM flavor symmetries.

Finally we come to $\beta_{q}$. Before we can determine this, we must 
first estimate the size of the dimensionless coefficient $c_q$ that 
parametrizes the bulk mass term. This is given by
\begin{equation}
c_{q,u} \sim \frac{\Lambda_{\text{IR}}}{k} \sim 4.9 \; .
\label{eq:c-NDA}
\end{equation}
If the mass term $c_q$ was of order its NDA size, then we would have 
$\sqrt{N}\beta_q \sim \sqrt{N}\alpha_q \sim 2.5$. However, in order to generate a realistic spectrum of fermion masses, it is necessary to take values of 
$c_q$ close to 1/2, significantly below its NDA value. It follows that 
the estimate of the coupling to the GW scalar is modified to
 \begin{equation} 
 \beta_q \sim \frac{2.5}{\sqrt{N}} \frac{c_q^{\rm NDA}}{c_q} \; , 
 \end{equation} 
 where $c_q^{\rm NDA}$ is given in Eq.~\eqref{eq:c-NDA}.

\bibliographystyle{JHEP}
\bibliography{thebigbibfile}

\end{document}